\documentclass[useAMS,usenatbib,deluxetable]{mn2e}

\usepackage{epsfig, natbib, graphicx, color,amsmath, cancel, icomma, bm, amssymb, deluxetable}

\input epsf 
    
\newcommand{\photoz}{photo$z$}
\newcommand{\zlambda}{z_{\lambda}}
\newcommand{\Pcen}{P_{\mathrm cen}}

\usepackage{color}

\newcommand{\hMpc}{h^{-1}\mbox{Mpc}}

\newcommand{\bmm}[1]{\mathbf{#1}}

\newcommand{\Nobs}{N_{obs}}

\newcommand{\avg}[1]{\left\langle #1 \right\rangle}

\newcommand{\lkhd}{\mbox{$\cal{L}$}}

\newcommand{\bC}{\bmm{C}}

\newcommand{\be}{\begin{equation}}
\newcommand{\ee}{\end{equation}}
\newcommand{\bea}{\begin{eqnarray}}
\newcommand{\eea}{\end{eqnarray}}

\newcommand{\kms}{\mathrm{km/s}}

\newcommand{\erf}{\mathrm{erf}}

\newcommand{\redmapper}{redMaPPer}
\newcommand{\pmem}{p_{\mathrm{mem}}}

\newcommand{\Nred}{N_{\mathrm{red}}}
\newcommand{\Nblue}{N_{\mathrm{blue}}}

\newcommand{\prm}{p_{\mathrm{RM}}}
\newcommand{\Ncut}{N_{\mathrm{cut}}}
\newcommand{\Nmem}{N_{\mathrm{mem}}}
\newcommand{\zcen}{z_{\mathrm{cen}}}

\newcommand{\rhored}{\rho_{\mathrm{red}}}

\newcommand{\pred}{p_{\mathrm{red}}}
\newcommand{\fred}{f_{\mathrm{red}}}
\newcommand{\rhotot}{\rho_{\mathrm{tot}}}

\newcommand{\chiobs}{\chi_{\mathrm{obs}}}
\newcommand{\chis}{\chi_{\mathrm{s}}}

\newcommand{\epsilonB}{\epsilon_{\mathrm{blue}}}
\newcommand{\epsilonchi}{\epsilon_{\chi^2}}
\newcommand{\fblue}{f_{\mathrm{blue}}}
\newcommand{\Ncorr}{N_{\mathrm{corr}}}

\newcommand{\Ntot}{N_\mathrm{tot}}
\newcommand{\rspec}{r_\mathrm{spec}}
\newcommand{\chitest}{\chi_\mathrm{test}}
\newcommand{\chiref}{\chi_\mathrm{ref}}

\newcommand{\half}{\frac{1}{2}}

\newcommand{\sref}{s_\mathrm{0}}

\title[Distinguishing Cluster Members with Photometric Data]{redMaPPer IV: Photometric Membership Identification of Cluster Galaxies with 1\% Precision}

\author[Rozo et al.]{E. Rozo$^{1,2,3}$, E. S. Rykoff$^{1,2}$, M. Becker$^{1,2}$, R. M. Reddick$^{1,2,4}$, R. H. Wechsler$^{1,2,4}$, 
 \\
$^{1}${Kavli Institute for Particle Astrophysics and Cosmology, P.O. Box 2450, Stanford, CA 94305}\\
$^{2}${SLAC National Accelerator Laboratory, 2575 Sand Hill Road,
  Menlo Park, CA 94025}\\
$^{3}${Department of Physics, University of Arizona, 1118 E 4th St, Tucson, AZ 85721}\\
$^{4}${Department of Physics, Stanford University, 382 Via Pueblo
  Mall, Stanford, CA 94305}
}

\begin{document}

\maketitle

\label{firstpage}

\begin{abstract}
  In order to study the galaxy population of galaxy clusters with
  photometric data one must be able to accurately discriminate between
  cluster members and non-members.  The \redmapper\ cluster finding
  algorithm treats this problem probabilistically.  Here, we utilize
  SDSS and GAMA spectroscopic membership rates to validate the \redmapper\ membership
  probability estimates for clusters with $z\in[0.1,0.3]$.  We find
  small --- but correctable --- biases, sourced by three different
  systematics.  The first two were expected a priori, namely blue
  cluster galaxies and correlated structure along the line of sight.
  The third systematic is new: the \redmapper\ template fitting
  exhibits a non-trivial dependence on photometric noise, which biases
  the original \redmapper\ probabilities when utilizing noisy data.
  After correcting for these effects, we find exquisite agreement
  ($\approx 1\%$) between the photometric probability estimates and
  the spectroscopic membership rates, demonstrating that we can
  robustly recover cluster membership estimates from photometric data
  alone.  As a byproduct of our analysis we find that on average unavoidable projection
  effects from correlated structure contribute $\approx 6\%$ of the richness of a \redmapper\
  galaxy cluster. 
  This work also marks the 
  second public release of the SDSS \redmapper\ cluster catalog.
\end{abstract}


\section{Introduction}

Galaxy clusters are not only powerful cosmological probes \citep[][]{henryetal09, vikhlininetal09, mantzetal10a, rozoetal10a, 
clercetal12b, bensonetal13, hasselfieldetal13, pXX}, they are also useful as galaxy evolution laboratories.  
Of particular interest are the relation between the mass of a cluster and the stellar mass and/or luminosity of 
its central galaxy \citep[e.g.][]{sheldonetal04,linmohr04,whileyetal08,moreetal09,moreetal11,pipinoetal11,skibbaetal11,edwardsetal12,kravtsovetal14a}, 
the satellite population \citep{skibbaetal07,yangetal09b,wetzelwhite10,budzynskietal12,nierenbergetal12,ruizetal13}, or 
both \citep{linetal04,mandelbaumetal06,hansenetal09,yangetal09,watsonconroy13,budzynskietal14}.  
Indeed, these relations are key predictions of semi-analytic models of galaxy formation \citep{liuetal10,quilistrujillo12},
of hydrodynamic simulations that aim to resolve galaxy formation \citep{kravtsovetal05,weinbergetal08,feldmannetal10,mccarthyetal10,mccarthyetal11,
martizzietal12,lebrunetal13,ragone-figueroaetal13,planellesetal13},
and of the popular sub-halo abundance matching scheme \citep[][]{conroyetal06,guoetal10,behroozietal10,reddicketal13b,kravtsov13,hearinetal13}.
Similarly, a comparison of the baryon budget in galaxy clusters
to the cosmic mean can provide valuable clues about the role of feedback and galaxy formation on the star formation efficiency
as a function of halo mass \citep{gonzalezetal07,giodinietal09,andreon10,laganaetal11,neisteinetal11,leauthaudetal12,linetal12,gonzalezetal13}.
Irrespective of the specific question being asked, any observational study that addresses these questions
must be able to robustly discriminate between galaxy cluster
members and unassociated galaxies along the line of sight.

Here, we investigate the ability of the \redmapper\ cluster finding algorithm \citep[][hereafter Paper I]{rykoffetal14} --- a new photometric algorithm
specifically optimized for multi-band photometric surveys like the Sloan Digital Sky Survey (SDSS), the Dark Energy Survey (DES), 
and the Large Synoptic Survey Telescope (LSST) --- to distinguish between cluster and non-cluster galaxies.   Specifically, 
\redmapper\ assigns a red membership probability $\prm$ to every galaxy in a cluster field.  
This probability is photometrically estimated, and includes
some assumptions that are known to be incorrect: \redmapper\ ignores both the existence of blue cluster
galaxies and of correlated structure along the line of sight. 
The goal of this paper is to investigate the impact these systematics have on the
\redmapper\ red membership probabilities.  

In order to test the \redmapper\ probabilities we rely on spectroscopic data from the SDSS \citep[SDSS DR10,][]{dr10} and 
from the Galaxy and Mass Assembly survey \citep[GAMA,][]{gama}.
Specifically, given a bin of spectroscopic galaxies with fixed photometric membership probability $\prm$, we empirically 
determine the fraction of red galaxies that are spectroscopic cluster members, and compare this fraction to $\prm$. 
As a by product of this analysis, we are also able to place a tight constraint on the impact of projections
from correlated structure on \redmapper\ cluster richness.  We emphasize that spectroscopic membership rates may not 
be trivially related to halo membership rates, and that this relation must necessarily depend on the halo definition being 
adopted \citep[e.g.][]{bivianoetal06, cohnetal07, serradiaferio13}.  Our work is exclusively concerned with the ``translation'' between photometric
to spectroscopic membership rates.  

The layout of the papers is as follows: section~\ref{sec:data} describes the data sets we use.  Section~\ref{sec:red_selection}
describes how we estimate the spectroscopic membership rates for red galaxies.  Section~\ref{sec:test} presents the
result of our spectroscopic membership rates before and after accounting for the biases in the \redmapper\ membership
probability estimates, which are also detailed in this section.  Section~\ref{sec:summary} summarizes our results and presents
a brief discussion.  Appendix~\ref{app:chibias} describes in detail our characterization of the photometric noise bias in the
\redmapper\ $\chi^2$ estimates.

We note that this work relies on the \redmapper\ v5.10 cluster catalog,
an updated version of the \redmapper\ v5.2 cluster catalog presented in
in Paper I.   Appendix \ref{app:updates} summarizes the changes and updates to the \redmapper\ catalog
relative to Paper I.  With this work we make the new \redmapper\ cluster catalog publicly available
at {\tt http://risa.stanford.edu/redmapper}.

%%%%%%%%%%%%%%%%%%%%%%%%%%%%%%%%%%%%%%%%
%%%%%%%%%%%%%%%%%%%%%%%%%%%%%%%%%%%%%%%%
%%%%%%%%%%%%%%%%%%%%%%%%%%%%%%%%%%%%%%%%
%%%%%%%%%%%%%%%%%%%%%%%%%%%%%%%%%%%%%%%%
%%%%%%%%%%%%%%%%%%%%%%%%%%%%%%%%%%%%%%%%
%%%%%%%%%%%%%%%%%%%%%%%%%%%%%%%%%%%%%%%%
%%%%%%%%%%%%%%%%%%%%%%%%%%%%%%%%%%%%%%%%
%%%%%%%%%%%%%%%%%%%%%%%%%%%%%%%%%%%%%%%%
%%%%%%%%%%%%%%%%%%%%%%%%%%%%%%%%%%%%%%%%
%%%%%%%%%%%%%%%%%%%%%%%%%%%%%%%%%%%%%%%%
%%%%%%%%%%%%%%%%%%%%%%%%%%%%%%%%%%%%%%%%
%%%%%%%%%%%%%%%%%%%%%%%%%%%%%%%%%%%%%%%%

\section{Data}
\label{sec:data}

\subsection{redMaPPer}

\redmapper\ is a red-sequence photometric cluster finding algorithm
that has been applied to SDSS DR8 photometric data \citep{dr8}.  This
galaxy catalog contains $\approx 14,000\ \deg^2$ of imaging, which we
reduce to $\approx 10,000\ \deg^2$ of contiguous high quality
observations using the mask from the Baryon Acoustic
Oscillation Survey (BOSS) \citep{dawsonetal13}.  \redmapper\ uses the
5-band ($ugriz$) data available for every galaxy, and imposes a
limiting magnitude of $i<21.0$.  For further details on the various
cuts and data handling employed by \redmapper, we refer the reader to
Paper I.  As mentioned above, the catalog employed in this work is an
updated version from that presented in Paper I, with the relevant
updates summarized in Appendix \ref{app:updates}.

Additionally, we have run the \redmapper{} algorithm on SDSS Stripe 82
(S82) coadd data~\citep{annisetal11}.  This catalog consists of
$275\,\mathrm{deg}^2$ of $ugriz$ coadded imaging over the equatorial
stripe that is roughly 2 magnitudes deeper than the single-pass SDSS
data used for the DR8 catalog.  Due to the higher redshift cutoff of
the \redmapper{} catalog in S82 data (we are roughly volume limited to
$z<0.7$), most of our member galaxies are $u$-band dropouts.  This,
combined with the challenges of accurate calibration of the $u$-band
data, led us to only use the $griz$ catalogs for input into
\redmapper.  We note that this catalog was used exclusively for demonstrating
the existence of photometric noise bias in our membership probability estimates,
as discussed in detail in Appendix~\ref{app:chibias}.  The stripe 82 catalog
is not being made public at this time.

To create the input galaxy catalog for \redmapper, we use similar flag
cuts as those used for DR8, described in Paper I.  In addition, we
clean all galaxies that have magnitude errors that are gross outliers
for typical galaxies at their observed magnitude.  As in Paper I,
total magnitudes are determined from $i$-band {\tt CMODEL\_MAG} and
colors from $griz$ {\tt MODEL\_MAG}.  To account for differences in
photometric calibration and survey depth, the \redmapper{}
red-sequence calibration is performed in the S82 data as usual,
irrespective of the DR8 data.

For reference, we briefly summarize the most salient features of the
\redmapper\ algorithm.  \redmapper\ is a matched filter algorithm.
The most important filter characterizes the color of red-sequence
galaxies as a function of redshift, which is self-calibrated by
relying on clusters with spectroscopic redshifts.  Having calibrated
the filters describing the red-sequence of galaxy clusters as a
function of redshift (amplitude, slope, and scatter), we use this
information to tag each galaxy in the vicinity of a galaxy cluster
with the probability $\prm$ of being a red cluster galaxy.  The
richness $\lambda$ is defined as the sum of the membership
probabilities over all galaxies,
\be
\lambda = \sum \prm. \label{eq:constraint}
\ee 

Here, unless otherwise specified, we will restrict our analysis of the DR8 \redmapper\ clusters to systems in the redshift range $z \in [0.1,0.3]$
with at least 20 galaxy counts.\footnote{The difference between richness and galaxy counts is that richness estimates accounts for cluster
masking and survey depth.  If a cluster is not masked at all, and the survey is sufficiently deep to detect all cluster galaxies brighter than $0.2L_*$,
then the richness is equal to the galaxy counts.} 
This results in $\approx 7000$ galaxy clusters spread over $\approx 10^4\ \deg^2$.   The lower redshift limit reflects the lowest
redshift at which \redmapper\ is expected to be properly calibrated, while the high redshift cutoff ensures that 
the \redmapper\ catalog is volume limited over the redshift range analyzed.  
The corresponding S82 \redmapper\ catalog contains nearly 2000 clusters above our selection threshold of 20 galaxy
counts, and the catalog is volume limited over the redshift range $z\in [0.1, 0.7]$.

%%%%%%%%%%%%%%%%%%%%%%%%%%%%%%%%%%%%%%%%
%%%%%%%%%%%%%%%%%%%%%%%%%%%%%%%%%%%%%%%%
%%%%%%%%%%%%%%%%%%%%%%%%%%%%%%%%%%%%%%%%
%%%%%%%%%%%%%%%%%%%%%%%%%%%%%%%%%%%%%%%%
%%%%%%%%%%%%%%%%%%%%%%%%%%%%%%%%%%%%%%%%
%%%%%%%%%%%%%%%%%%%%%%%%%%%%%%%%%%%%%%%%
%%%%%%%%%%%%%%%%%%%%%%%%%%%%%%%%%%%%%%%%
%%%%%%%%%%%%%%%%%%%%%%%%%%%%%%%%%%%%%%%%
%%%%%%%%%%%%%%%%%%%%%%%%%%%%%%%%%%%%%%%%
%%%%%%%%%%%%%%%%%%%%%%%%%%%%%%%%%%%%%%%%
%%%%%%%%%%%%%%%%%%%%%%%%%%%%%%%%%%%%%%%%
%%%%%%%%%%%%%%%%%%%%%%%%%%%%%%%%%%%%%%%%

\subsection{Spectroscopic Data}

Our spectroscopic membership test relies on two distinct spectroscopic data sets.
The first is SDSS DR10 \citep{dr10}.  DR10 combines
all available spectroscopy from the SDSS through DR9 and new spectra
acquired as part of the BOSS experiment.  The total number of galaxy
spectra in DR10 is 927,844, comprising a magnitude-limited sample \citep[main sample][]{straussetal02},
the SDSS Luminous Red Galaxy sample \citep[LRG][]{eisensteinetal01}, 
and the BOSS targets, which includes an approximately constant stellar mass sample
at high redshifts (CMASS) and a low redshift red galaxy sample \citep{dawsonetal13}.

The second spectroscopic data set used is from the Galaxy And Mass
Assembly survey (GAMA).  GAMA is a magnitude-limited ($r<19.8$)
spectroscopic survey of $\approx 300,000$ galaxies over $\approx 290\
\deg^2$, carried out using the AAOmega multi-object spectrograph on
the Anglo-Australian Telescope \citep{gama}.  Here, we utilize GAMA
data through the second data
release\footnote{http://www.gama-survey.org/dr2/} and apply a quality
flag $\geq 3$ cut, which includes $\approx 70,000$ galaxy redshifts
over $\approx 48\ \deg^2$, all of which overlaps with the footprint of
the \redmapper\ cluster catalog.

In order to assign spectra from either SDSS or GAMA to the \redmapper\ photometric
member list of galaxies we rely on positional matching
using a 1" angular aperture.    To test the validity of this procedure as well as the robustness
of the SDSS and GAMA redshifts, we perform this angular matching between the
SDSS DR 10 and GAMA spectroscopic catalogs, and then study the distribution
of the redshift difference $\Delta z = |z_\mathrm{SDSS} - z_\mathrm{GAMA}|$.

%%%%%%%%%%%%%%%%%%%%%%%%%%%%%%%%%%%%%%%%
%%%%%%%%%%%%%%%%%%%%%%%%%%%%%%%%%%%%%%%%

\begin{figure}
\hspace{-12pt} \includegraphics[width=90mm]{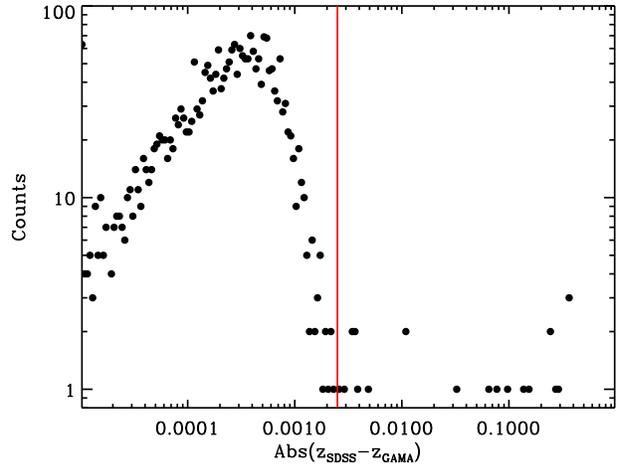}
\caption{Number of galaxies within logarithmically spaced bins of 
$\Delta z = |z_\mathrm{SDSS}-z_\mathrm{GAMA}|$.  Redshifts offsets
to the right of the red vertical line are considered catastrophic failures,
and represent $0.9\%$ of the objects.
}
\label{fig:zcomp}
\end{figure}

%%%%%%%%%%%%%%%%%%%%%%%%%%%%%%%%%%%%%%%%
%%%%%%%%%%%%%%%%%%%%%%%%%%%%%%%%%%%%%%%%

Figure~\ref{fig:zcomp} shows the distribution of $\Delta z$ in logarithmically
spaced bins.  We see that there is a tail of discrepant redshifts starting at
$\Delta z \approx 0.0025$, marked by the vertical red line.  The total fraction
of objects with a redshift difference larger than this is $0.9\%$, which we
adopt as our estimate of the spectroscopic redshift failure rate, which contributes
to our systematic uncertainty for the spectroscopic membership rate of photometric
cluster members.

%%%%%%%%%%%%%%%%%%%%%%%%%%%%%%%%%%%%%%%%
%%%%%%%%%%%%%%%%%%%%%%%%%%%%%%%%%%%%%%%%
%%%%%%%%%%%%%%%%%%%%%%%%%%%%%%%%%%%%%%%%
%%%%%%%%%%%%%%%%%%%%%%%%%%%%%%%%%%%%%%%%
%%%%%%%%%%%%%%%%%%%%%%%%%%%%%%%%%%%%%%%%
%%%%%%%%%%%%%%%%%%%%%%%%%%%%%%%%%%%%%%%%
%%%%%%%%%%%%%%%%%%%%%%%%%%%%%%%%%%%%%%%%
%%%%%%%%%%%%%%%%%%%%%%%%%%%%%%%%%%%%%%%%
%%%%%%%%%%%%%%%%%%%%%%%%%%%%%%%%%%%%%%%%
%%%%%%%%%%%%%%%%%%%%%%%%%%%%%%%%%%%%%%%%
%%%%%%%%%%%%%%%%%%%%%%%%%%%%%%%%%%%%%%%%
%%%%%%%%%%%%%%%%%%%%%%%%%%%%%%%%%%%%%%%%

\section{Selecting Red Spectroscopic Cluster Members}
\label{sec:red_selection}

We wish to test the \redmapper\ membership probabilities by comparing
to spectroscopic membership rates.  That is, if we select \redmapper\ galaxies
with membership probability $90\%$, and the \redmapper\ membership probabilities
are correct, then 90\% of the selected galaxies ought to be spectroscopic
cluster members.  The first step in performing this test then is to describe how
we estimate the spectroscopic membership rate of a set of galaxies.

\subsection{Red Galaxy Spectroscopic Membership Rates}
\label{sec:specmem}

Given a galaxy cluster with velocity dispersion $\sigma_v$,
we define a spectroscopic member to be a galaxy within a projected
aperture $R_\lambda$ (see below) of a cluster, and whose velocity
along the line of sight satisfies
\be
|v| \leq \Nmem \sigma_v \label{eq:vmem}
\ee
where $\Nmem$ is a fiducial threshold, and $\sigma_v(\lambda,z)$ is the velocity
dispersion of a cluster of richness $\lambda$ at redshift $z$.  The radius $R_\lambda$
is the radius used by \redmapper\ to estimate the cluster richness, and is related
to the cluster richness via
\be
R_\lambda = 1.0\ \hMpc \left( \frac{\lambda}{100} \right)^{0.2}.
\ee

One obvious problem with defining cluster membership via equation \ref{eq:vmem} is that cluster
membership depends on the adopted membership cut $\Nmem$.  
We can account for this difficulty by assuming that the line of sight velocity distribution of galaxies
is Gaussian.  Specifically, if
$\Nobs(\Nmem)$ is the number of spectroscopic galaxies obtained using a velocity cut $|v| \leq \Nmem \sigma_v$,
the completeness-corrected number of members is
\be
\Ntot = \Nobs/\erf( \Nmem/\sqrt{2} ). \label{eq:Ncorrection}
\ee
If the velocity distribution were exactly Gaussian, the above definition would result
in spectroscopic membership estimates that are independent of the adopted $\Nmem$
cut.    In practice, we find that our spectroscopic
membership rates are robust to changes in $\Nmem$ over the range $\Nmem \in [1,2.5]$ at the $1\%$ level,
which we adopt as the associated systematic uncertainty.  We choose $\Nmem=2.0$ as our fiducial membership
threshold.  We add in quadrature the systematic uncertainty in the spectroscopic membership rate due to our choice
of $\Nmem$ to the spectroscopic redshift failure rate to arrive at a net systematic uncertainty of $1.3\%$. 
For reference, at the pivot point of our data, a $2\sigma_v$ cut in redshift corresponds to a $\approx 10\ \hMpc$
separation along the line-of-sight.

Now, as was noted in the introduction, the \redmapper\ probability $\prm$
is meant to describe the probability that a galaxy is a {\it red} cluster galaxy.
Since blue cluster galaxies exist, we do not expect the \redmapper\ membership
probability $\prm$ to agree with the total fraction of spectroscopic cluster members,
as some of those members will not be red galaxies.
We must therefore limit ourselves to red galaxies when computing the spectroscopic membership rate.

This is easily done: let $\pred$ be the probability of a galaxy being a red galaxy.  Given $N$ galaxies
each with a red probability $\pred$, the total number of red galaxies is simply the sum total of the
red membership probabilities.  To obtain the total number of red cluster members, one simply
restricts the sum to galaxies with $|v| \leq \Nmem \sigma_v$, exactly as before.  If the total number
of galaxies is $N$, the completeness-corrected red spectroscopic membership rate is simply
\be
\rspec = \frac{1}{\erf(\Nmem/\sqrt{2})} \frac{1}{N}\sum_{|v|\leq \Nmem\sigma_v} \pred.
\label{eq:rspec}
\ee

Equation \ref{eq:rspec} is the fundamental equation we use to determine the spectroscopic membership
rate that is to be compared to the \redmapper\ membership probabilities.  
In the following two sections, we describe how we estimate $\sigma_v$ and $\pred$ for every cluster
and galaxy respectively.

%%%%%%%%%%%%%%%%%%%%%%%%%%%%%%%%%%%%%%%%
%%%%%%%%%%%%%%%%%%%%%%%%%%%%%%%%%%%%%%%%
%%%%%%%%%%%%%%%%%%%%%%%%%%%%%%%%%%%%%%%%
%%%%%%%%%%%%%%%%%%%%%%%%%%%%%%%%%%%%%%%%
%%%%%%%%%%%%%%%%%%%%%%%%%%%%%%%%%%%%%%%%
%%%%%%%%%%%%%%%%%%%%%%%%%%%%%%%%%%%%%%%%
%%%%%%%%%%%%%%%%%%%%%%%%%%%%%%%%%%%%%%%%
%%%%%%%%%%%%%%%%%%%%%%%%%%%%%%%%%%%%%%%%
%%%%%%%%%%%%%%%%%%%%%%%%%%%%%%%%%%%%%%%%
%%%%%%%%%%%%%%%%%%%%%%%%%%%%%%%%%%%%%%%%
%%%%%%%%%%%%%%%%%%%%%%%%%%%%%%%%%%%%%%%%
%%%%%%%%%%%%%%%%%%%%%%%%%%%%%%%%%%%%%%%%

\subsection{Estimating Velocity Dispersions}
\label{sec:veldisp}

In the previous section we described how to estimate the cluster membership rate
for galaxies in clusters of known velocity dispersion.  Unfortunately, 
our spectroscopic data set is such that the typical number of galaxy spectra per cluster
is low, so it is impossible for us to provide robust estimates of the velocity dispersion 
of individual clusters.  Instead, we first calibrate the scaling relation between
cluster velocity dispersion and cluster richness, and then use the velocity dispersions
estimated from this scaling relation to determine cluster membership.

We calibrate the $\sigma_v$--$\lambda$ scaling relation as follows.  First, 
we select all \redmapper\ clusters whose central galaxy has 
a spectroscopic redshift, and then search for spectroscopic galaxies within the cluster
radius $R_\lambda$ of each such cluster, irrespective of whether or not the galaxies
are included in the cluster member list.  In order to maximize our statistics,
we employ both SDSS DR10 and GAMA spectroscopy.
For each central--satellite par, we compute the velocity offset
\be
v = c \frac{z_\mathrm{sat}-\zcen}{1+\zcen}
\ee
where $c$ is the speed of light.

%%%%%%%%%%%%%%%%%%%%%%%%%%%%%%%%%%%%%%%%
%%%%%%%%%%%%%%%%%%%%%%%%%%%%%%%%%%%%%%%%

\begin{figure}
\hspace{-12pt} \includegraphics[width=90mm]{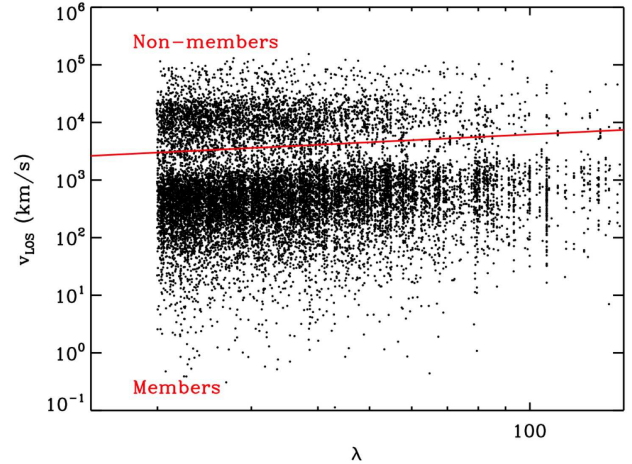}
\caption{Line-of-sight velocities of spectroscopic central--satellite pairs
of \redmapper\ galaxy clusters.  The red line shows the initial split into
spectroscopic members and non-cluster members used to initialize
the calibration of the velocity dispersion of the clusters.
}
\label{fig:vlos}
\end{figure}

%%%%%%%%%%%%%%%%%%%%%%%%%%%%%%%%%%%%%%%%
%%%%%%%%%%%%%%%%%%%%%%%%%%%%%%%%%%%%%%%%

Figure \ref{fig:vlos} shows the velocity offset $v$ of each central--satellite pair in the
sample.  There are two obvious populations: a sample
with low velocity offsets ($|v| \lesssim 2000\ \kms)$, and a population of high velocity
offsets.  We perform an initial selection criteria for spectroscopic members via
\be
|v| \leq (3000\ \kms) (\lambda/20)^{0.45}
\ee
shown in Figure \ref{fig:vlos} as a red line.  

We model the velocity distribution of cluster satellite galaxies
as a Gaussian of mean $\avg{v}=0$ with and a velocity dispersion $\sigma_v$
that is richness and redshift dependent,
\be
\sigma_v(\lambda,\zcen) = \sigma_p \left( \frac{1+\zcen}{1+z_p}\right)^\beta \left( \frac{\lambda}{\lambda_p} \right)^\alpha.
\ee
In the above expression, $\lambda_p$ is a pivot point chosen a priori to be the median richness 
of the cluster sample ($\lambda_p=33.336$), and $z_p$ is the median cluster redshift of all
velocity pairs ($z_p=0.171$).  $\sigma_p$ is the velocity dispersion of a cluster of richness $\lambda_p$ 
at the pivot redshift, and $\alpha$ and $\beta$ characterize the dependence of the velocity dispersion 
with cluster richness and redshift respectively.  

We model the contribution of non-cluster members as a uniform background of spectroscopic galaxies.   
The full likelihood for any one galaxy is 
\be
\lkhd_i = p G(v_i) + (1-p)\frac{1}{2v_\mathrm{max}}
\ee
where $v_\mathrm{max}$ is the maximum velocity cut used to define the sample of candidate members.
Here, $G(v_i)$ is a Gaussian of mean zero and velocity dispersion $\sigma_v(\lambda,\zcen)$.  The model
parameters are $p$, $\sigma_p$, $\alpha$, and $\beta$, and the total likelihood is obtained by multiplying
the individual likelihoods for every central--satellite pair,
\be
\lkhd = \prod_i \lkhd_i.
\ee
Our best fit parameters are obtained by maximizing this likelihood.

To find our best fit parameters, we initialize our fits by setting $\alpha=0.45$ based on a ``by eye'' fit (e.g. Figure~\ref{fig:vlos}).
We further set $\beta=0$ (i.e. ignore redshift evolution), and do a Gaussian fit to the velocity histogram obtained by stacking
all clusters, irrespective of richness.   The resulting width is adopted as the initial value for the amplitude $\sigma_p$.  
We then refit the data using our likelihood method after applying a selection cut
\be
|v| \leq \Ncut\sigma_v
\label{eq:cut}
\ee
with $\Ncut=5.0$, so that the constant background is most representative of the areas immediately adjacent to the cluster member
population.  We refit, and the procedure is iterated until convergence.
We arrive at
\bea
\lambda_p & = & 33.336 \\
z_p & = & 0.171 \\
\sigma_p & = & (618.1 \pm 6.0) \ \kms \\
\alpha & = & 0.435 \pm 0.020 \\
\beta & = & 0.54 \pm 0.19 \\
p & = & 0.9163 \pm 0.0042
\eea
All errors are estimated using $10^3$ bootstrap resamplings of the velocity pairs, and are nearly uncorrelated.

We have further repeated the above procedure setting  $\Ncut=6.0$.  The difference
in the recovered parameters is completely insignificant, except for the background
amplitude parameter $p$ which must, of course, vary. 
We note that this robustness was only achieved when modeling the background.
In particular, utilizing $\Ncut \in [2,3]$ with no background modeling results in systematic
errors at the $\sim 8\%$ level, with the results being clearly dependent on $\Ncut$.

%%%%%%%%%%%%%%%%%%%%%%%%%%%%%%%%%%%%%%%%
%%%%%%%%%%%%%%%%%%%%%%%%%%%%%%%%%%%%%%%%

\begin{figure}
\hspace{-12pt} \includegraphics[width=90mm]{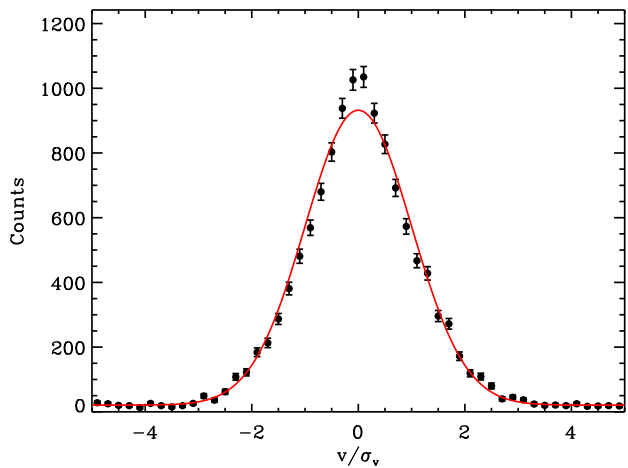}
\caption{Distribution of the line-of-sight velocity $v$, normalized by the expected velocity dispersion $\sigma_v$
of the cluster, for all central--satellite pairs in our sample.  The red curve is our best fit model 
(i.e. a gaussian of zero mean and width unity, and a flat background of non cluster galaxies).  
The amplitude is set by demanding that the total number of galaxy counts in the model curve exactly
match the total number of counts in the data.
}
\label{fig:vhist}
\end{figure}

%%%%%%%%%%%%%%%%%%%%%%%%%%%%%%%%%%%%%%%%
%%%%%%%%%%%%%%%%%%%%%%%%%%%%%%%%%%%%%%%%

Figure \ref{fig:vhist} compares the distribution of normalized
line-of-sight velocity offsets $v/\sigma_v$ for central--satellite
pairs in our sample.  The red line is a Gaussian of zero mean and unit
variance, to which we have added the appropriate background model as
per our best fit.  The curve has been normalized so that the integral
is equal to the total number of velocity pairs in the plot.  Our model
is {\it not} a good fit to the data ($\chi^2/dof=96.0/26$ for
$v/\sigma_v \in [-3,3]$); the data is somewhat more sharply peaked
than the unit Gaussian at $v\approx 0$.  Consequently, it is not
appropriate to use our likelihood to estimate the confidence interval
of our best fit parameters, thereby explaining our reliance of
bootstrap resampling.  We emphasize, however, that the goal of this
work is {\it not} to perform a detailed calibration of the
$\sigma_v$--$\lambda$ relation: our main interest is to be able to
identify cluster members from spectroscopic data, and our model
suffices for this task.  In particular, as noted earlier, the
systematic error in our spectroscopic membership rates due to using a
Gaussian model as per equation~\ref{eq:Ncorrection} is only 1\%.  A
detailed analysis of the $\sigma_v$--$\lambda$ relation will be
presented in a future paper.

Finally, we consider how uncertainty in our scaling relation impacts
our spectroscopic membership rates.  Varying the amplitude of the
relation by its allotted error, we find that the recovered membership
rates vary by $\pm 0.4\%$.  Added in quadrature to the above error
estimate we arrive at a total systematic error of $1.4\%$.  Varying
the remaining scaling relation parameters results in a negligible
($\leq 0.1\%$) change in the membership rates.  This easily
understood: changing the richness or redshift slopes will increase the
membership rate on one side of the pivot point while simultaneously
decreasing the membership rate on the opposite side of the pivot
point.  Because our pivot point is chosen to be at the median cluster
richness/redshift, these two perturbations very nearly cancel each
other.

%%%%%%%%%%%%%%%%%%%%%%%%%%%%%%%%%%%%%%%%
%%%%%%%%%%%%%%%%%%%%%%%%%%%%%%%%%%%%%%%%
%%%%%%%%%%%%%%%%%%%%%%%%%%%%%%%%%%%%%%%%
%%%%%%%%%%%%%%%%%%%%%%%%%%%%%%%%%%%%%%%%
%%%%%%%%%%%%%%%%%%%%%%%%%%%%%%%%%%%%%%%%
%%%%%%%%%%%%%%%%%%%%%%%%%%%%%%%%%%%%%%%%
%%%%%%%%%%%%%%%%%%%%%%%%%%%%%%%%%%%%%%%%
%%%%%%%%%%%%%%%%%%%%%%%%%%%%%%%%%%%%%%%%
%%%%%%%%%%%%%%%%%%%%%%%%%%%%%%%%%%%%%%%%
%%%%%%%%%%%%%%%%%%%%%%%%%%%%%%%%%%%%%%%%
%%%%%%%%%%%%%%%%%%%%%%%%%%%%%%%%%%%%%%%%
%%%%%%%%%%%%%%%%%%%%%%%%%%%%%%%%%%%%%%%%

\subsection{When is a Galaxy Red?}
\label{sec:pred}

We determine whether a galaxy is red or not by evaluating the goodness
of fit of the \redmapper\ red-sequence template.  Roughly speaking, a
galaxy is red if the red sequence template fit provides a good fit to
its photometry (i.e. has a low $\chi^2$).  In practice, however, our
treatment of red galaxies is more sophisticated as detailed below.

\begin{figure*}
\hspace{-12pt} \includegraphics[width=90mm]{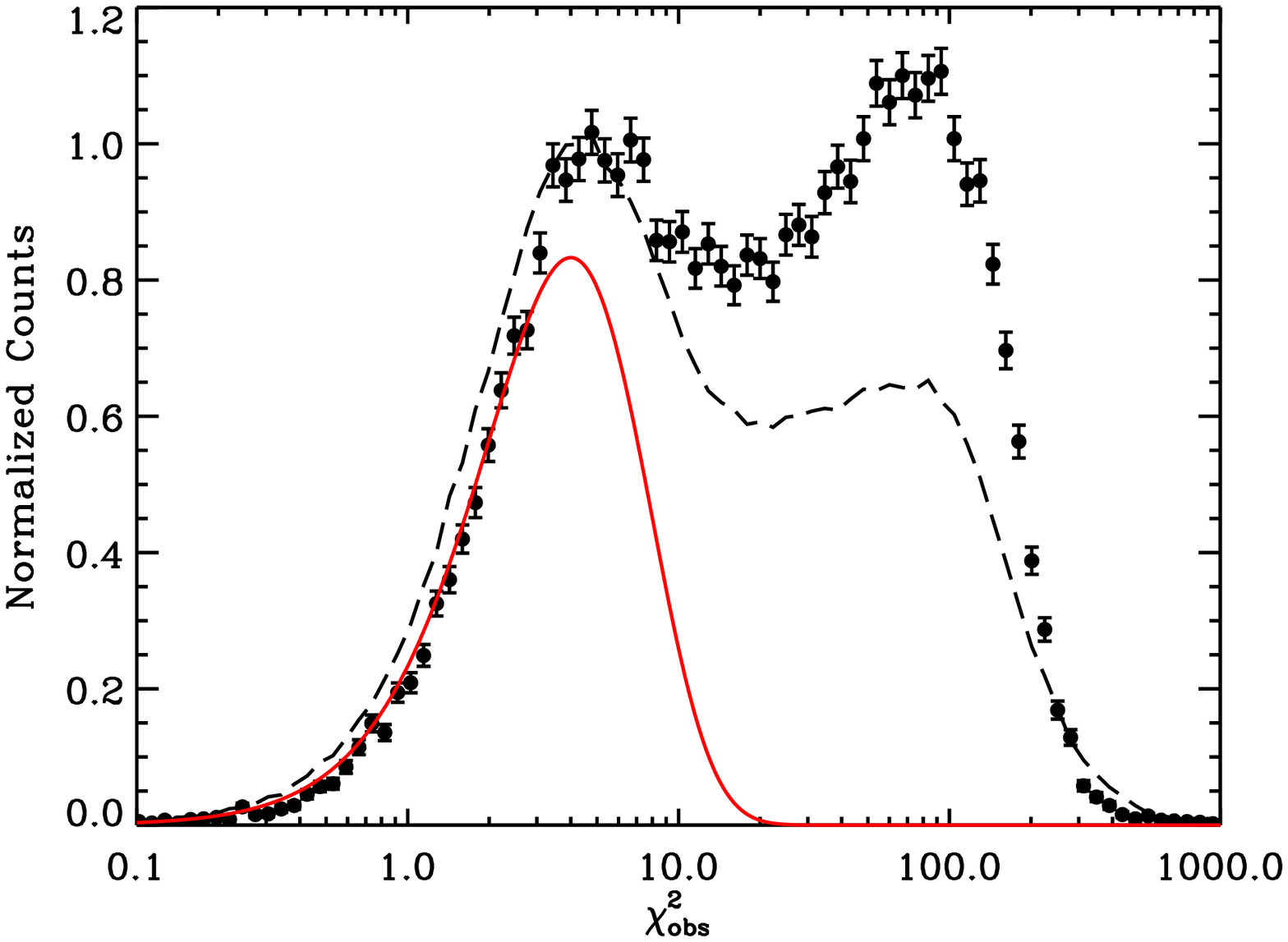}
\hspace{-12pt} \includegraphics[width=90mm]{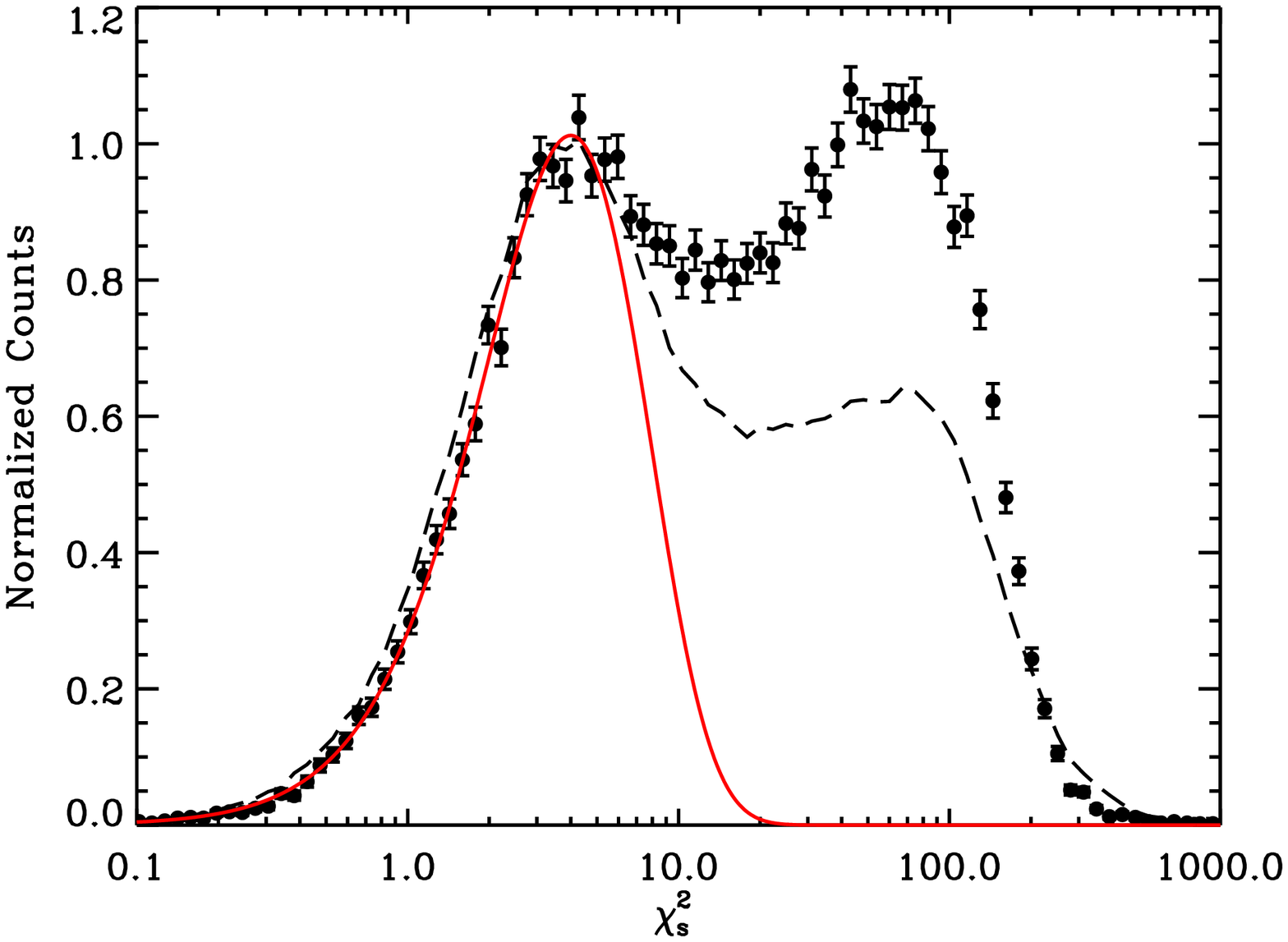}
\caption{{\bf Left panel:} Distribution of the goodness-of-fit statistic $\chiobs^2$ for all GAMA galaxies (points with error bars).
$\chiobs^2$ is the goodness of fit of each galaxy's photometry to the \redmapper\ red-sequence template
at the galaxy's spectroscopic redshift.  The GAMA data is normalized to unity at its left-peak value, 
estimated by fitting a cubic function to the data points around the peak.   The dashed line is the corresponding
distribution for all SDSS DR10 spectroscopic galaxies.  Also shown for
reference is a $\chi^2$ distribution with four degrees 
of freedom (red line), normalized so as to have the same integral as the GAMA data over the range $\chiobs^2 \in [0,4]$.
{\bf Right panel:} As left panel, but using the rescaled $\chi^2$ values that correct for the effects of photometric
noise bias.  
}
\label{fig:chihist}
\end{figure*}

The left panel of Figure \ref{fig:chihist} shows the distribution of $\chi^2$ --- the goodness of fit of the \redmapper\ red-sequence
template to the galaxy photometry --- for all GAMA galaxies with $z\in[0.1,0.3]$, as evaluated at each galaxy's spectroscopic redshift
(the red-sequence template is redshift dependent, see Paper I for details).
Also shown is the $\chi^2$ distribution obtained using SDSS spectra only (dashed curve).  In both cases, we have normalized
the distributions to equal unity at their corresponding red peaks.

Also shown with a red curve is a $\chi^2$ distribution with 4 degrees of freedom (there are 4 colors), which we expect to be a good descriptor 
of the red wings of the $\chi^2$ distributions above.  To fit the distribution of $\chi^2$ values with a $\chi^2$ distribution 
(allowing for an overall normalization factor only), it is imperative that the fit be performed over a $\chi^2$ range over which
there is no contamination by non-red galaxies.   Here, we have normalized the $\chi^2$ distribution by demanding that
the integral of the $\chi^2$ distribution over the range $\chi^2 \in [0,4]$ agree with that of the empirical distribution.  

%%%%%%%%%%%%%%%%%%%%%%%%%%%%%%%%%%%%%%%%
%%%%%%%%%%%%%%%%%%%%%%%%%%%%%%%%%%%%%%%%

%%%%%%%%%%%%%%%%%%%%%%%%%%%%%%%%%%%%%%%%
%%%%%%%%%%%%%%%%%%%%%%%%%%%%%%%%%%%%%%%%

If our covariance matrices were all exactly correct, the red wings of the SDSS and GAMA galaxies should
fall directly on top of each other, and they would agree with the red curve.   This is clearly not the case.
As we show in Appendix \ref{app:chibias}, our $\chi^2$ values suffer from photometric noise biases,
and it is this bias which is responsible for the differences seen in the left panel of Figure~\ref{fig:chihist}.

In Appendix \ref{app:chibias} we demonstrate the photometric noise bias in our $\chi^2$ values can be removed
by rescaling our $\chi^2$ values via
\be
\chis^2 = \exp(-s)\chiobs^2 \label{eq:sdef}.
\ee
where $s$ accounts for the bias in our observed $\chi^2$.  The bias $s$ is unique to each galaxy, and depends
on each galaxy's photometric errors.  For details, see Appendix \ref{app:chibias}.

The right panel in Figure~\ref{fig:chihist} shows the distribution of the rescaled $\chis^2$ values for SDSS and GAMA,
as well as our reference $\chi^2$ distribution.  As before, the $\chi^2$ distribution is normalized based on the integrated
counts of the empirical distribution for $\chi^2\in[0,4]$.  We see that the agreement between the various distributions has
been dramatically improved.   For the rest of this section, we rely exclusively on the rescaled 
$\chis$ values for every galaxy whenever we refer to $\chi^2$.  Nevertheless, we keep the subscript `s' to make the rescaling 
explicit throughout.

The agreement between a $\chi^2$ distribution and the distribution of $\chis^2$ values for spectroscopic
galaxies seen in the right panel of Figure~\ref{fig:chihist} enables us to define a probability $\pred$ for any
given galaxy to be a red galaxy.  
Specifically, let $\rhored$ and $\rhotot$
be the distribution of $\chi^2$ values for red galaxies and all galaxies respectively.  The distribution $\rhored$ is defined
to be a $\chi^2$ distribution with 4 degrees of freedom, but we demand that the integral of this distribution over the
region $\chi^2 \in [0,4]$ match the integral of the empirical distribution of $\chis^2$ values over the same range.
The probability that a galaxy of a given $\chi^2$ is a red galaxy is simply
\be
\pred = \frac{\rhored}{\rhotot}.
\label{eq:pred}
\ee
%

%%%%%%%%%%%%%%%%%%%%%%%%%%%%%%%%%%%%%%%%
%%%%%%%%%%%%%%%%%%%%%%%%%%%%%%%%%%%%%%%%
%%%%%%%%%%%%%%%%%%%%%%%%%%%%%%%%%%%%%%%%
%%%%%%%%%%%%%%%%%%%%%%%%%%%%%%%%%%%%%%%%
%%%%%%%%%%%%%%%%%%%%%%%%%%%%%%%%%%%%%%%%
%%%%%%%%%%%%%%%%%%%%%%%%%%%%%%%%%%%%%%%%
%%%%%%%%%%%%%%%%%%%%%%%%%%%%%%%%%%%%%%%%
%%%%%%%%%%%%%%%%%%%%%%%%%%%%%%%%%%%%%%%%
%%%%%%%%%%%%%%%%%%%%%%%%%%%%%%%%%%%%%%%%
%%%%%%%%%%%%%%%%%%%%%%%%%%%%%%%%%%%%%%%%
%%%%%%%%%%%%%%%%%%%%%%%%%%%%%%%%%%%%%%%%
%%%%%%%%%%%%%%%%%%%%%%%%%%%%%%%%%%%%%%%%

\subsection{Environmental and Luminosity Dependences of the Red Fraction}
\label{sec:dependences}

The probability $\pred$ depends
of both environment and galaxy luminosity; after all, bright galaxies tend to be red, and 
cluster galaxies tend to be red.   We investigate how the distribution of $\chis^2$ values
depends on both galaxy luminosity and environment by considering 4 different galaxy 
subsamples.  The different samples are
\begin{itemize}
\item All GAMA galaxies with $z\in[0.1,0.3]$.
\item All GAMA galaxies in $z\in[0.1,0.3]$ which are also spectroscopic cluster members ($2\sigma_v$ cut).
\item All bright GAMA galaxies with $z\in[0.1,0.3]$.  
\item All dim GAMA galaxies with $z\in[0.1,0.3]$.
\end{itemize}
Note that we have restricted ourselves to GAMA spectroscopy (which has a magnitude-limited
target selection) in order to avoid any biases due to color selection in SDSS targeting.  
To define bright and dim galaxies, we rank order
the GAMA galaxies by $m-m_*(z)$, where $m_*(z)$ is the apparent luminosity of an $L_*$ galaxy
utilized by the \redmapper\ algorithm.  The sample is then split into thirds, with the bright sample being
the brightest third, and the dim sample being the dimmest third.

\begin{figure*}
\hspace{-12pt} \includegraphics[width=90mm]{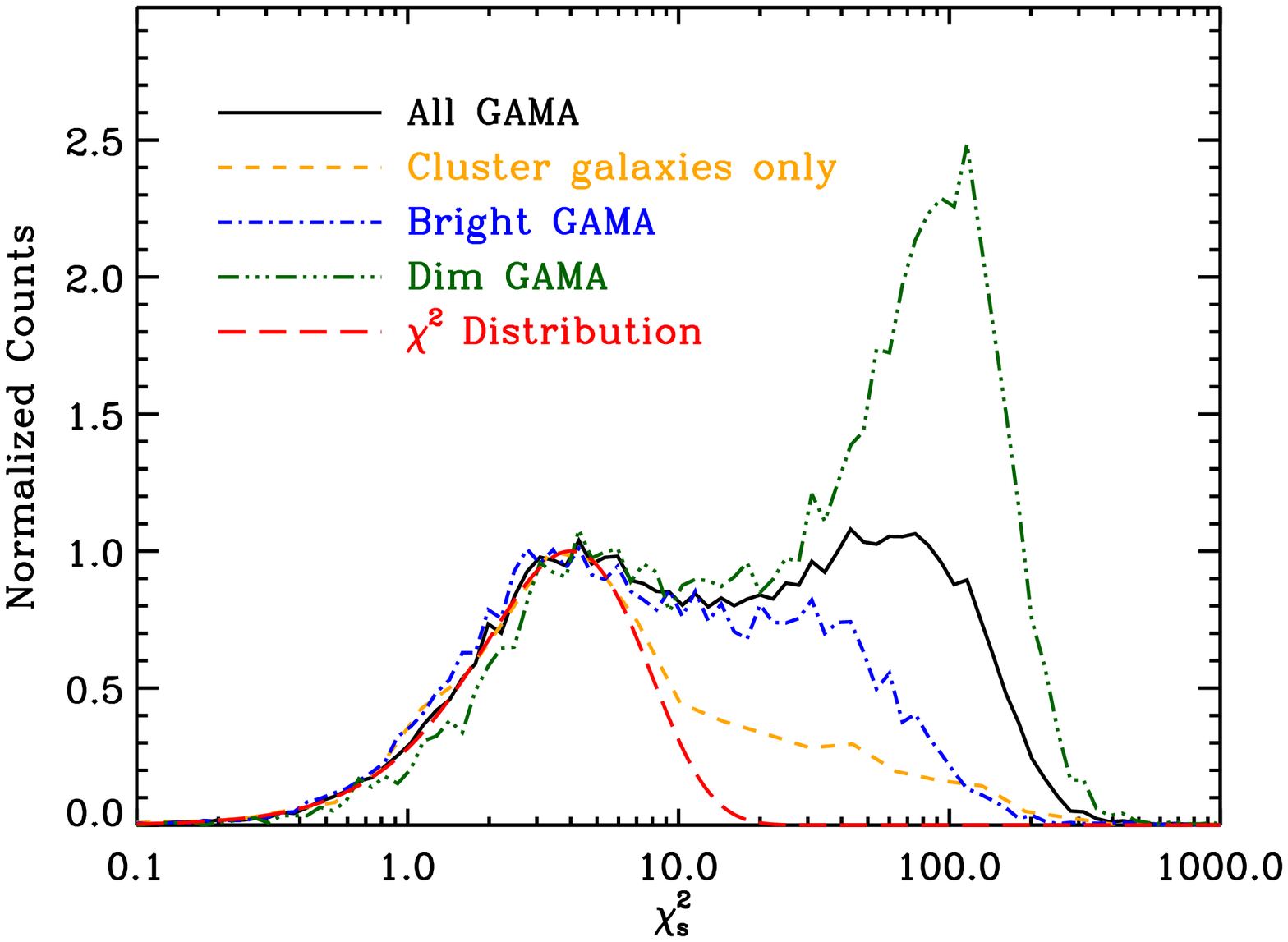}
\hspace{-12pt} \includegraphics[width=90mm]{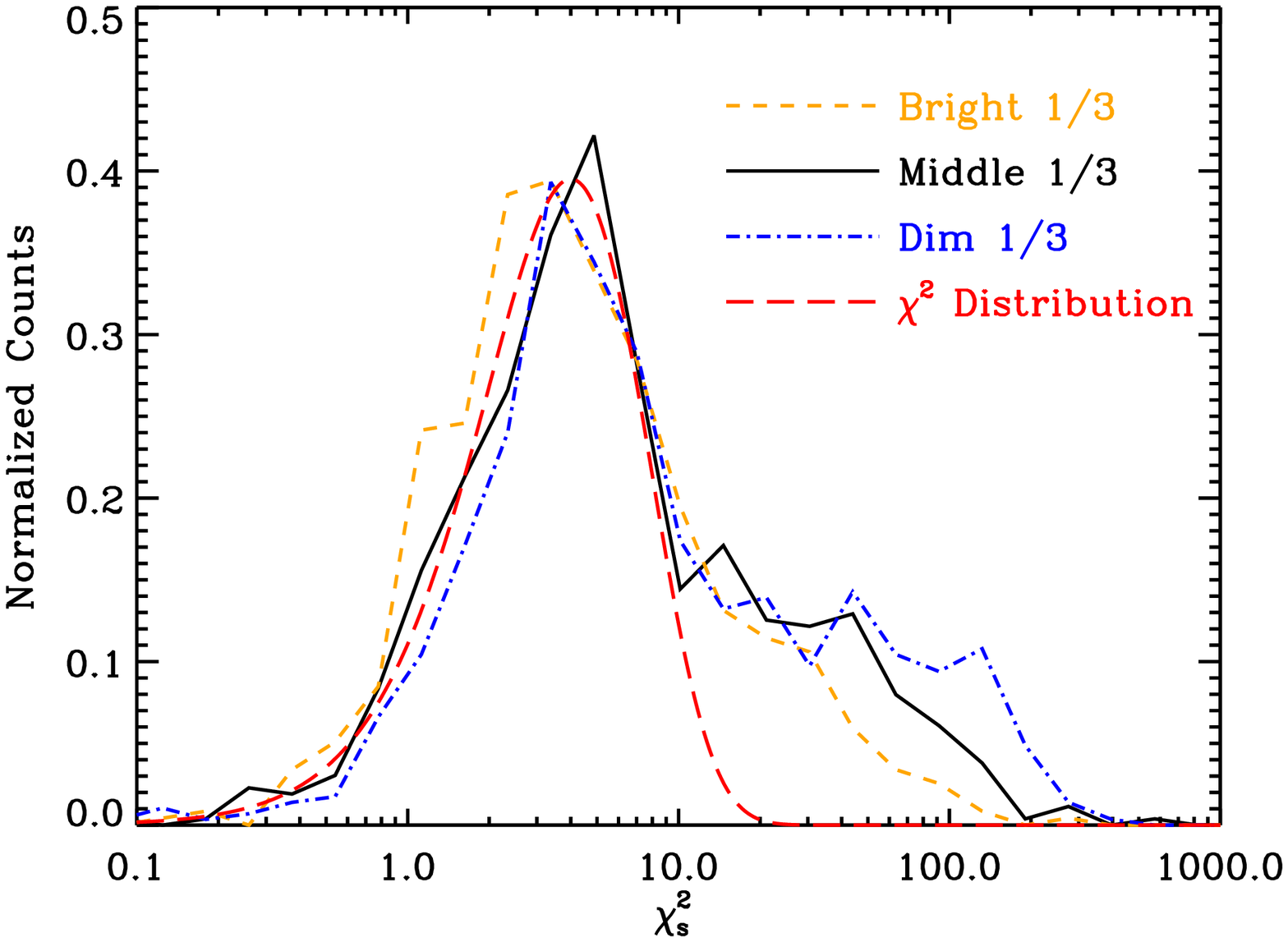}
\caption{{\bf Left panel:} Comparison of the $\chis^2$ distribution of galaxies selected in a variety of ways, as labelled (see text
for detailed descriptions).  All distributions are normalized to unity at their left-most peak, and are well described by
a $\chi^2$ distribution with 4 degrees of freedom (red dashed line). {\bf Right panel: } 
Comparison of the $\chis^2$ distribution of {\it cluster} galaxies for three different luminosity bins,
as labelled. The red long-dash curve is a $\chi^2$ distribution with four degrees of freedom, normalized to have
the same integral over the region $\chi^2\in[0,6]$ as the full data set.
All empirical distributions are normalized to have the same integral over all space.
}
\label{fig:dependences}
\end{figure*}

The left panel of Figure~\ref{fig:dependences} shows the $\chis^2$ distribution of each of our galaxy
samples, as labelled.   
It is immediately apparent that irrespective of any selection effects, the red wing of the $\chis^2$ distribution
is well described by a $\chi^2$ distribution with 4 degrees of freedom.  
Additionally, there is an obvious luminosity and environmental dependence of the $\chis^2$ distributions: 
dim galaxies have a much larger ratio of blue-to-red galaxies than bright galaxies, and 
cluster galaxies are very strongly preferentially red.
Evidently, we must account for both the impact of environment and galaxy luminosity on the red fraction.

%%%%%%%%%%%%%%%%%%%%%%%%%%%%%%%%%%%%%%%%
%%%%%%%%%%%%%%%%%%%%%%%%%%%%%%%%%%%%%%%%

%%%%%%%%%%%%%%%%%%%%%%%%%%%%%%%%%%%%%%%%
%%%%%%%%%%%%%%%%%%%%%%%%%%%%%%%%%%%%%%%%

The right panel of Figure~\ref{fig:dependences} shows the $\chis^2$ distribution of cluster galaxies in three luminosity
bins, each chosen to contain 1/3 of the available GAMA spectra.  
Due to the comparatively low
number of spectra available for this analysis, we have decreased the \redmapper\ richness threshold
to $\lambda=10$ in order to make this plot.  All histograms were normalized to an integral of unity,
which makes evident the rather surprising result that the $\chi^2$ distribution of cluster galaxies is 
roughly luminosity independent.   

\begin{figure}
\hspace{-12pt} \includegraphics[width=90mm]{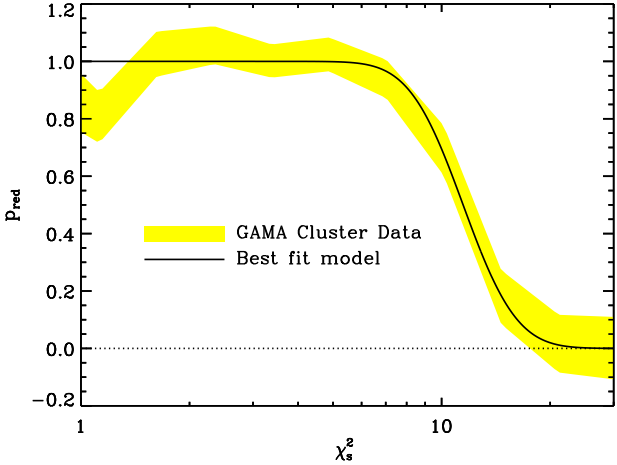}
\caption{Probability $\pred=\rhored/\rhotot$ (i.e. the red fraction) as a function of $\chis^2$.  
The bands marks the empirical 68\% confidence regions, while the black line shows
our best fit model.
}
\label{fig:pred}
\end{figure}

In light of the above results, we combine all GAMA spectroscopic cluster galaxies into a single bin, and ignore
any possible luminosity dependence of the $\chi^2$ distribution.  We then fit the resulting distribution with a
$\chi^2$ distribution with 4 degrees of freedom by demanding that 
the integral of the model agree with the data over the range $\chi^2 \in[0,6]$, with $\chi^2\approx 6$
being roughly the largest $\chi^2$ value for which $\pred=1$.   This defines the distribution $\rhored$
which we use in equation~\ref{eq:pred} to estimate $\pred$.  
Our resulting $\pred(\chis^2)$ function is shown in Figure~\ref{fig:pred}.

We fit the probability $\pred(\chis)$ via
\be
\pred(\chis) = \frac{1}{2}\left[ 1- \erf\left( \frac{\ln( \chis/\chiref)}{\sqrt{2}\sigma} \right)\right]
\ee
where $\chiref$ and $\sigma$ are parameters to be fit for.  We find
\bea
\ln \chiref & = & 2.44 \pm 0.08  \\
\sigma & = & 0.28 \pm 0.11 
\eea
with the two parameters being nearly uncorrelated.  Our best fit model for $\pred(\chis)$ 
is also shown in Figure~\ref{fig:pred} as a solid black line.  
In all subsequent work, unless otherwise specified when we need to evaluate $\pred(\chis)$,
we will rely on our best fit model to do so.  In particular, we use this best fit model
to compute the red spectroscopic membership rates as per equation \ref{eq:rspec}.

%%%%%%%%%%%%%%%%%%%%%%%%%%%%%%%%%%%%%%%%
%%%%%%%%%%%%%%%%%%%%%%%%%%%%%%%%%%%%%%%%

%%%%%%%%%%%%%%%%%%%%%%%%%%%%%%%%%%%%%%%%
%%%%%%%%%%%%%%%%%%%%%%%%%%%%%%%%%%%%%%%%

We note that the uncertainty associated with whether a galaxy is red or not implies a corresponding
systematic uncertainty in the red spectroscopic membership rates.  This uncertainty is easily dominated
by the error  in $\ln \chiref$, which induces an error of $\pm 1.6\%$. 
When added in quadrature to our previous estimate of the systematic uncertainty, we arrive at a net
systematic error of $2.1\%$ for our red spectroscopic membership rates.

%%%%%%%%%%%%%%%%%%%%%%%%%%%%%%%%%%%%%%%%
%%%%%%%%%%%%%%%%%%%%%%%%%%%%%%%%%%%%%%%%
%%%%%%%%%%%%%%%%%%%%%%%%%%%%%%%%%%%%%%%%
%%%%%%%%%%%%%%%%%%%%%%%%%%%%%%%%%%%%%%%%
%%%%%%%%%%%%%%%%%%%%%%%%%%%%%%%%%%%%%%%%
%%%%%%%%%%%%%%%%%%%%%%%%%%%%%%%%%%%%%%%%
%%%%%%%%%%%%%%%%%%%%%%%%%%%%%%%%%%%%%%%%
%%%%%%%%%%%%%%%%%%%%%%%%%%%%%%%%%%%%%%%%
%%%%%%%%%%%%%%%%%%%%%%%%%%%%%%%%%%%%%%%%
%%%%%%%%%%%%%%%%%%%%%%%%%%%%%%%%%%%%%%%%
%%%%%%%%%%%%%%%%%%%%%%%%%%%%%%%%%%%%%%%%
%%%%%%%%%%%%%%%%%%%%%%%%%%%%%%%%%%%%%%%%

\section{The Spectroscopic Membership Test}
\label{sec:test}

\subsection{Testing the \redmapper\ Membership Probabilities}
\label{sec:spectest}

We compare the \redmapper\ membership probabilities $\prm$ to the
red spectroscopic membership rate as estimated via equation \ref{eq:rspec}.
Specifically, we collect \redmapper\ cluster member galaxies into membership
probability bins, and compute the mean membership probability of each bin.
This mean probability is then compared to the measured spectroscopic membership
rate.  

The results of this comparison is shown in the left panel of Figure~\ref{fig:spectest}.
The agreement between the photometric probabilities and the spectroscopic membership
rates is reasonable, but there is an obvious bias: \redmapper\ systematically overestimates
the membership probabilities by $\approx 5\%$.  As we now demonstrate,
this bias is well understood and can be fully accounted for.

%%%%%%%%%%%%%%%%%%%%%%%%%%%%%%%%%%%%%%%%
%%%%%%%%%%%%%%%%%%%%%%%%%%%%%%%%%%%%%%%%
%%%%%%%%%%%%%%%%%%%%%%%%%%%%%%%%%%%%%%%%
%%%%%%%%%%%%%%%%%%%%%%%%%%%%%%%%%%%%%%%%
%%%%%%%%%%%%%%%%%%%%%%%%%%%%%%%%%%%%%%%%
%%%%%%%%%%%%%%%%%%%%%%%%%%%%%%%%%%%%%%%%
%%%%%%%%%%%%%%%%%%%%%%%%%%%%%%%%%%%%%%%%
%%%%%%%%%%%%%%%%%%%%%%%%%%%%%%%%%%%%%%%%
%%%%%%%%%%%%%%%%%%%%%%%%%%%%%%%%%%%%%%%%
%%%%%%%%%%%%%%%%%%%%%%%%%%%%%%%%%%%%%%%%
%%%%%%%%%%%%%%%%%%%%%%%%%%%%%%%%%%%%%%%%
%%%%%%%%%%%%%%%%%%%%%%%%%%%%%%%%%%%%%%%%

\subsection{Understanding the Biases in the \redmapper\ Probabilities}
\label{sec:understanding}

The bias in the \redmapper\ probability estimates are a combination of three separate
effects. Specifically,
\begin{enumerate}
\item Photometric noise bias in the \redmapper\ $\chi^2$ values.
\item \redmapper\ ignores correlated structure.
\item \redmapper\ ignores blue cluster galaxies.
\end{enumerate}

Consider first photometric noise biases in $\chi^2$.  Our de-biased $\chi^2$
estimates are given by equation \ref{eq:s}.  Assuming these rescaled $\chi^2$
values are distributed via a $\chi^2$ distribution with four degrees of freedom, which
we denote $\rho_0$, it follows that the distribution of the original $\chi^2$ values is
\be
\rho = \exp(-s)\rho_0( \exp(-s)\chiobs^2 ).
\ee
The membership probability is therefore
\bea
\pmem & = & \frac{\lambda \rho}{\lambda\rho + B} \label{eq:prm} \\
	& = & \frac{ \lambda \rho_0 (1+\epsilonchi) }{(\lambda\rho_0+B)\left(1+(\lambda \rho_0 \epsilonchi)/(\lambda\rho_0+B) \right) } \\
	& = & \prm \frac{ 1+\epsilonchi }{1+\prm \epsilonchi} \label{eq:filter}
\eea
where $B$ is the background, $\prm$ is the original \redmapper\ membership probability estimate, and
\be
\epsilonchi = \frac{\rho(\chiobs^2)}{\rho_0(\chiobs^2)} - 1.
\ee
Equation~\ref{eq:filter} allows us to correct the effects of photometric noise bias in $\chi^2$ on the \redmapper\ probability $\prm$.

We can perform a similar calculation for the impact of correlated structure.  Assuming that the correlated galaxy counts $\Ncorr$
is a constant fraction $c$ of the cluster richness, we find
\bea
\pmem & = & \prm \frac{1}{1+\prm c} \label{eq:corr}
\eea
Finally, \redmapper\ ignores the existence of blue galaxies.   Consequently, 
the true probability that a galaxy is a red cluster member is not $\prm$, but rather $\pmem$,
where
\bea
\pmem & = & \frac{\Nred}{\Nred+\Nblue+B} \\
	    & = & \frac{\prm}{1+\prm \epsilonB}. \label{eq:blue}
\eea
where
\be
\epsilonB = \frac{\fblue(\chi^2)}{1-\fblue(\chi^2)}
\ee
and $\fblue(\chi^2)=1-\pred$ is the blue fraction as a function of $\chi^2$.

There is one additional effect that must be properly accounted for: as we vary
the membership probability of galaxies, we must also vary the total cluster richness in concert,
since the two are related by the constraint equation \ref{eq:constraint}.   A shift in cluster richness
$\lambda=\lambda_0(1+\delta)$ will necessarily rescale all membership probabilities via
\be
\pmem = \prm \frac{1+\delta}{1+\prm \delta}
\label{eq:delta}
\ee
and vice versa.  Given the probability rescaling detailed above, we re-estimate the cluster richness by 
summing up the new probabilities.  This new richness estimate is used to compute the parameter $\delta$,
which is then used to rescale the membership probabilities as per equation~\ref{eq:delta}.  
The procedure is then iterated one more time.  We find that additional iterations perturb our
membership below the $0.5\%$ level, and are therefore negligible.

We note that while for pedagogical purposes we considered each perturbation in isolation, in practice we 
simultaneously consider the impact of all of the effects considered here.  We find that the \redmapper\
membership probabilities must be rescaled via
\be
\pmem = \prm \frac{ 1+ \nu}{1+\prm(\mu+\nu+\mu\nu)}
\label{eq:pmem}
\ee
where
\bea
\mu & = & \epsilonB+c \\
\nu & = & \delta+\epsilonchi+\delta\epsilonchi.
\eea
%

%%%%%%%%%%%%%%%%%%%%%%%%%%%%%%%%%%%%%%%%
%%%%%%%%%%%%%%%%%%%%%%%%%%%%%%%%%%%%%%%%
%%%%%%%%%%%%%%%%%%%%%%%%%%%%%%%%%%%%%%%%
%%%%%%%%%%%%%%%%%%%%%%%%%%%%%%%%%%%%%%%%
%%%%%%%%%%%%%%%%%%%%%%%%%%%%%%%%%%%%%%%%
%%%%%%%%%%%%%%%%%%%%%%%%%%%%%%%%%%%%%%%%
%%%%%%%%%%%%%%%%%%%%%%%%%%%%%%%%%%%%%%%%
%%%%%%%%%%%%%%%%%%%%%%%%%%%%%%%%%%%%%%%%
%%%%%%%%%%%%%%%%%%%%%%%%%%%%%%%%%%%%%%%%
%%%%%%%%%%%%%%%%%%%%%%%%%%%%%%%%%%%%%%%%
%%%%%%%%%%%%%%%%%%%%%%%%%%%%%%%%%%%%%%%%
%%%%%%%%%%%%%%%%%%%%%%%%%%%%%%%%%%%%%%%%

\subsection{Calibration of Projection Effects}

%%%%%%%%%%%%%%%%%%%%%%%%%%%%%%%%%%%%%%%%
%%%%%%%%%%%%%%%%%%%%%%%%%%%%%%%%%%%%%%%%

\begin{figure}
\hspace{-12pt} \includegraphics[width=90mm]{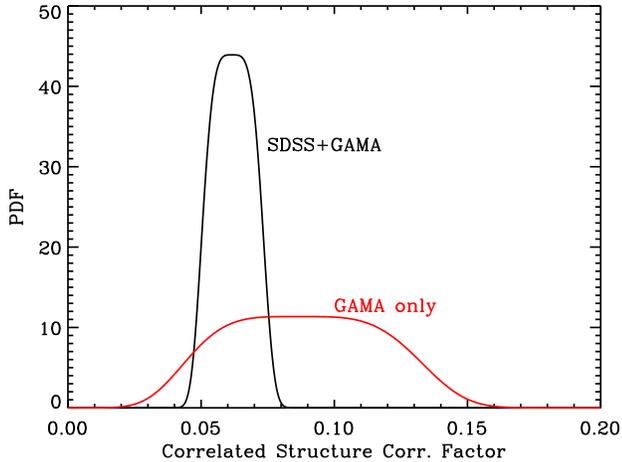}

\caption{
Posteriors on the mean fraction of red photometric cluster members selected
by the \redmapper\ algorithm that are not spectroscopic cluster members, but are
instead contributed by correlated structures along the line of sight.  
}
\label{fig:correlated}
\end{figure}

%%%%%%%%%%%%%%%%%%%%%%%%%%%%%%%%%%%%%%%%
%%%%%%%%%%%%%%%%%%%%%%%%%%%%%%%%%%%%%%%%

The parameter $c$ governing the impact of projection effects in \redmapper\ clusters is 
unknown a priori. 
We utilize our spectroscopic membership test in order to calibrate the parameter
$c$.
Specifically, given a value of $c$, we can use equation \ref{eq:pmem}
to rescale all of the membership probabilities for galaxies in the \redmapper\ cluster member
catalog, and compare these to the spectroscopic membership rates as per section~\ref{sec:spectest}.
This allows us to compute a goodness-of-fit statistic $\chitest^2$, where we use the subscript
``test'' to distinguish this $\chi^2$ value from the other occurrences of $\chi^2$ in this manuscript.  
We have then 
\be
\chitest^2 = \sum \frac{ (\fred^{(i)} - \pmem^{(i)})^2 }{\sigma_i^2}.
\ee
where the sum is over the membership probability bins. The error $\sigma_i$ is
given by
\be
\sigma_i = \frac{\sqrt{N_\mathrm{specmems}}}{N_i} 
\ee

We adopte a likelihood $\lkhd \propto \exp(-\chitest^2)$, and grid in the parameter $c$ to measure
the corresponding likelihood distribution, which we show in Figure~\ref{fig:correlated}. 
We fit for $c$ using both GAMA and SDSS+GAMA data sets, checking for consistency to
guard ourselves against biases introduced by the impact of color selection in the SDSS
spectroscopic targeting algorithm.  We find consistent results between the two data sets.
The 68\% confidence interval for the SDSS+GAMA data set is
$c=6.2\% \pm 0.8\%$ (see Figure~\ref{fig:correlated}).  
Finally, recall that the spectroscopic membership rates are themselves uncertain
at the $\approx \pm 2.1\%$ level.  Adding all of these quantities in quadrature we arrive 
at
\be
c = 6.2\% \pm 2.2\%.
\label{eq:cobs}
\ee
This is an important observational constraint on any future model of projection effects.
We again caution, however, that our spectroscopic membership rate is not equivalent to
a halo membership rate, and in particular the relation between these two is necessarily dependent
on the halo definition adopted, so the appropriate conversions must be undertaken when interpreting
our results within the context of a halo model.
The corresponding value for the parameter $\delta$ is $\delta=-0.094$. 
The dominant uncertainty in this analysis is the systematic error associated with our
ability to determine whether a galaxy is red or not --- i.e. the uncertainty in the probability
$\pred(\chis^2)$ --- followed closely by the systematic error in the spectroscopic membership
rate due to the non-Gaussian nature of the velocity distribution of galaxies in galaxy clusters.

%%%%%%%%%%%%%%%%%%%%%%%%%%%%%%%%%%%%%%%%
%%%%%%%%%%%%%%%%%%%%%%%%%%%%%%%%%%%%%%%%
%%%%%%%%%%%%%%%%%%%%%%%%%%%%%%%%%%%%%%%%
%%%%%%%%%%%%%%%%%%%%%%%%%%%%%%%%%%%%%%%%
%%%%%%%%%%%%%%%%%%%%%%%%%%%%%%%%%%%%%%%%
%%%%%%%%%%%%%%%%%%%%%%%%%%%%%%%%%%%%%%%%
%%%%%%%%%%%%%%%%%%%%%%%%%%%%%%%%%%%%%%%%
%%%%%%%%%%%%%%%%%%%%%%%%%%%%%%%%%%%%%%%%
%%%%%%%%%%%%%%%%%%%%%%%%%%%%%%%%%%%%%%%%
%%%%%%%%%%%%%%%%%%%%%%%%%%%%%%%%%%%%%%%%
%%%%%%%%%%%%%%%%%%%%%%%%%%%%%%%%%%%%%%%%
%%%%%%%%%%%%%%%%%%%%%%%%%%%%%%%%%%%%%%%%

\subsection{Testing the Rescaled Membership Probabilities}

\begin{figure*}
\hspace{-12pt} \includegraphics[width=90mm]{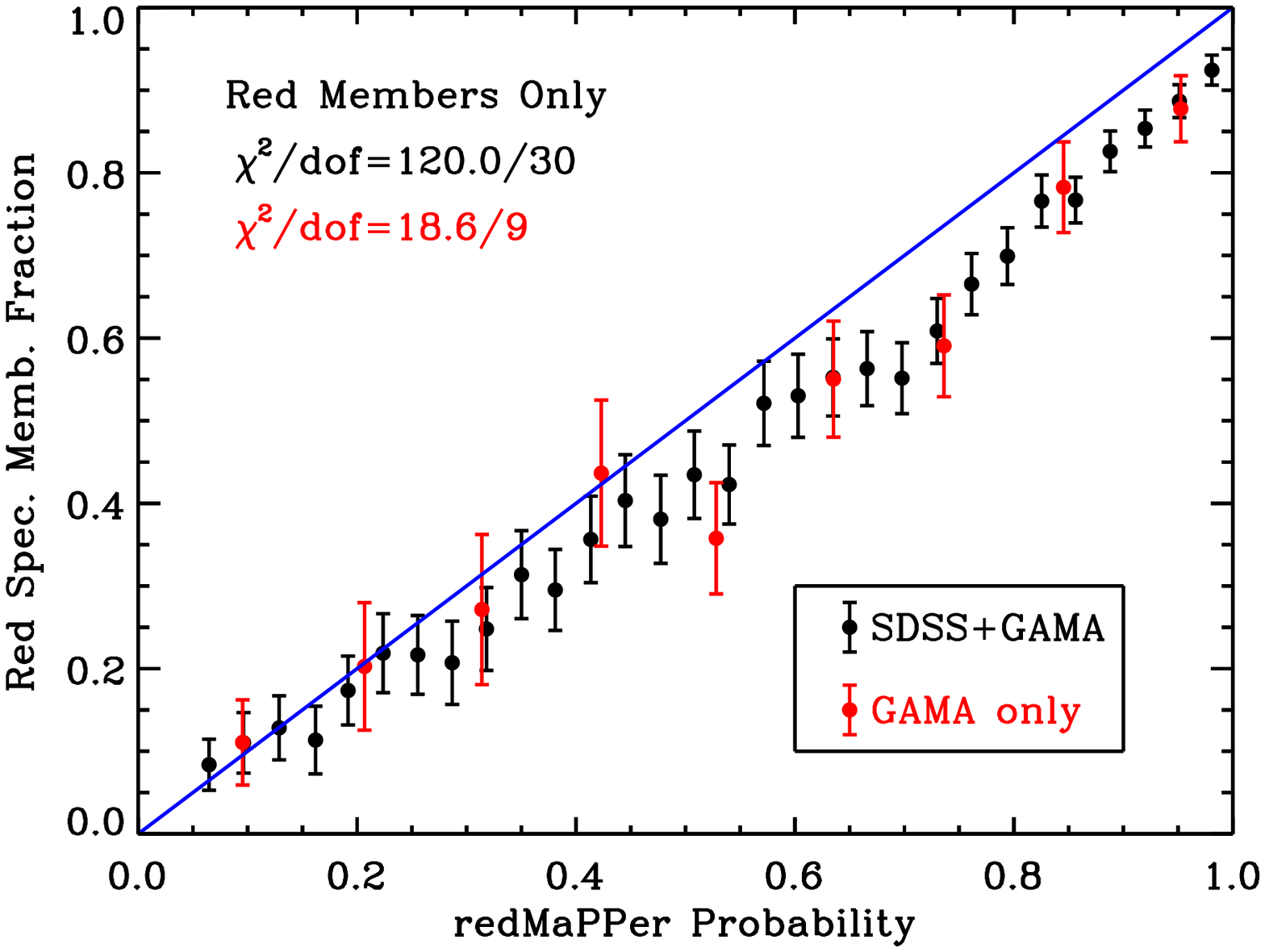}
\hspace{-12pt} \includegraphics[width=90mm]{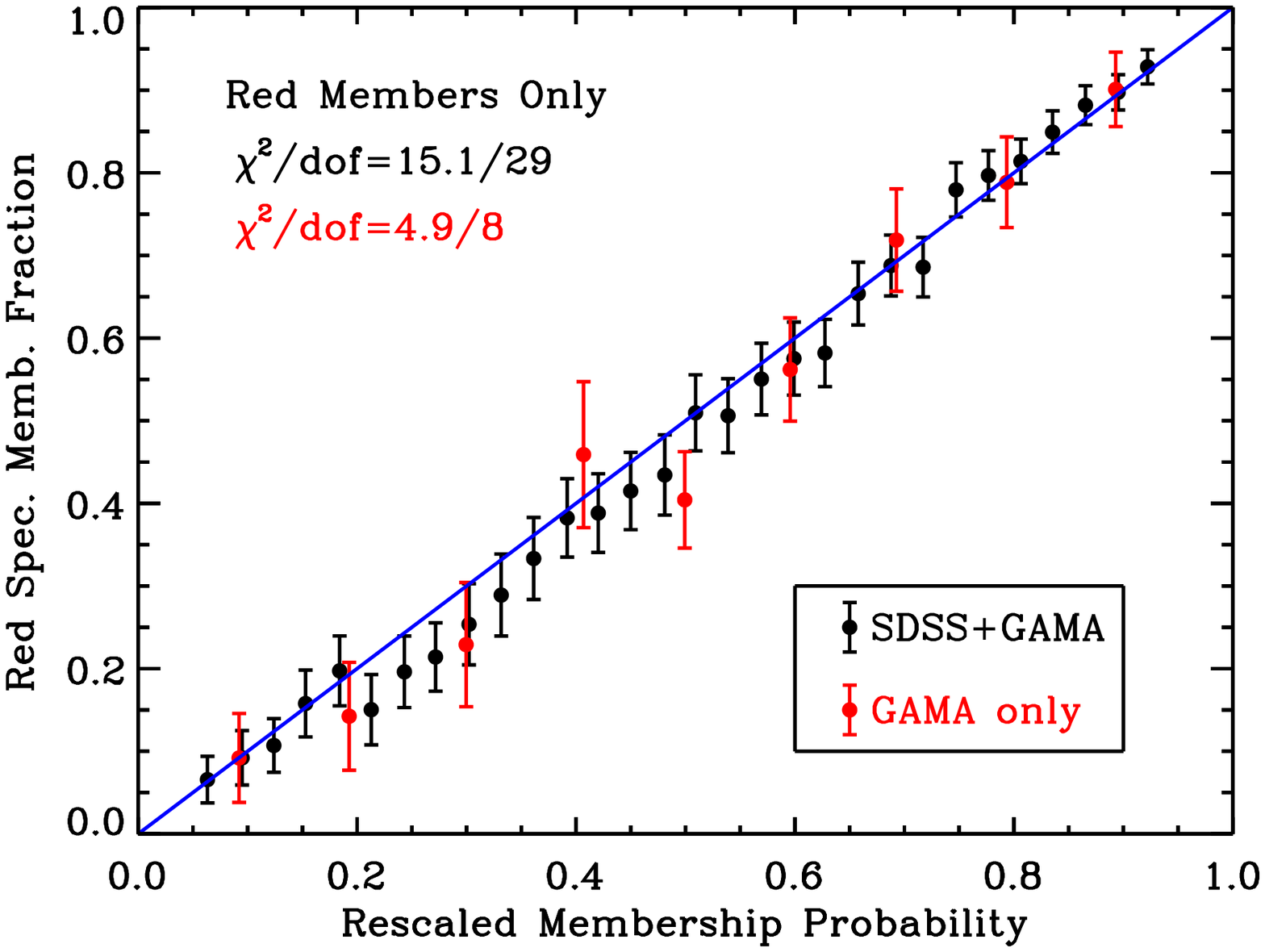}
\caption{
{\bf Left panel:}   Comparison of the \redmapper\ membership probabilities to the red
spectroscopic membership rate.  The observed biases are due to the combination of
three systematic effects: noise biases in the $\chi^2$ values used by \redmapper,
the fact that \redmapper\ ignores the existence of both cluster galaxies and correlated
structure.  {\bf Right panel: } As left panel, but after correcting for
the three aforementioned systematic effects.  The correction for projection
effects introduces a free parameters that is being fit to the data, as reflected in the
decrease in the degrees of freedom for the $\chi^2$ goodness-of-fit test.
}
\label{fig:spectest}
\end{figure*}

Figure~\ref{fig:spectest} compares the rescaled photometric membership probabilities $\pmem$ 
to the red spectroscopic membership rates.  
We find the rescaled membership probabilities provide somewhat too good a fit to the spectroscopic data
\be
\chi^2/dof = 15.1/29.
\ee
The probability of finding a $\chi^2$ larger than observed is $\approx 98\%$ ($2.3\sigma$).  It is likely that this
low $\chi^2$ reflects a failure of our statistical modeling.  For instance, at $\pmem\approx 0.3$,
there is a broad region where our rescaled probabilities appear to be biased somewhat low across
many nominally independent points, suggesting that the points are not in fact statistically independent.
Nevertheless, the agreement between the photometric and spectroscopic membership rates is remarkable.

%%%%%%%%%%%%%%%%%%%%%%%%%%%%%%%%%%%%%%%%
%%%%%%%%%%%%%%%%%%%%%%%%%%%%%%%%%%%%%%%%

%%%%%%%%%%%%%%%%%%%%%%%%%%%%%%%%%%%%%%%%

%%%%%%%%%%%%%%%%%%%%%%%%%%%%%%%%%%%%%%%%
%%%%%%%%%%%%%%%%%%%%%%%%%%%%%%%%%%%%%%%%
%%%%%%%%%%%%%%%%%%%%%%%%%%%%%%%%%%%%%%%%
%%%%%%%%%%%%%%%%%%%%%%%%%%%%%%%%%%%%%%%%
%%%%%%%%%%%%%%%%%%%%%%%%%%%%%%%%%%%%%%%%
%%%%%%%%%%%%%%%%%%%%%%%%%%%%%%%%%%%%%%%%
%%%%%%%%%%%%%%%%%%%%%%%%%%%%%%%%%%%%%%%%
%%%%%%%%%%%%%%%%%%%%%%%%%%%%%%%%%%%%%%%%
%%%%%%%%%%%%%%%%%%%%%%%%%%%%%%%%%%%%%%%%
%%%%%%%%%%%%%%%%%%%%%%%%%%%%%%%%%%%%%%%%
%%%%%%%%%%%%%%%%%%%%%%%%%%%%%%%%%%%%%%%%
%%%%%%%%%%%%%%%%%%%%%%%%%%%%%%%%%%%%%%%%

\section{Summary and Discussion}
\label{sec:summary}

We have studied whether the photometrically estimated \redmapper\
membership probabilities can be used to accurately determine whether
any given galaxy is a red cluster member or not.  The raw \redmapper\
probabilities are biased relative to the observed spectroscopic
membership rates, which is expected given that \redmapper\ explicitly
assumes that there are not blue galaxies in clusters, and that
clusters have no correlated structure, both assumptions that are
obviously incorrect a priori.

In addition, we have found that the \redmapper\ $\chi^2$ values suffer
from noise bias.  This bias is typically $\approx 25\%$ for DR8 data,
and could be thought of as a systematic bias in the estimate of the
covariance matrix describing the red-sequence.  Note that a $25\%$
bias in $\chi^2$ corresponds to a 12\% bias in the red-sequence
scatter, or $\approx 0.006$ mag.  While it is not clear to us what the
physical origin of this bias is, our work demonstrates that the noise
bias can be empirically characterized and accounted for (see
Appendix~\ref{app:chibias}).

Having identified the sources of bias in the membership probability
estimates from \redmapper, we corrected our probability estimates,
including a fit for the amount of average contribution from correlated
structure to the \redmapper\ cluster richness.  The corrected
membership probabilities are observed to be in excellent agreement
with the spectroscopic membership rate, with an overall systematic
uncertainty of 2.4\%.  For reference, the systematic floor due to
spectroscopic redshift failure rates is 0.9\%.  In other words, our
calibration enables studies of the galaxy population of galaxy
clusters from photometric data without incurring a significant
degradation in the quality of the data relative to a fully
spectroscopic data set.

Interestingly, as a byproduct of this analysis we were able to
constrain the average contribution to a cluster's richness due to
projected structure in the low redshift Universe, 
finding that, on average, 6.2\% of the richness
of a galaxy cluster is due to non-cluster galaxies.  This is an
important observational constraint that can be used to better
characterize the impact of projection effects on photometric cluster
samples, for instance within the context of cluster cosmology in the
DES or LSST.  As noted in the text, we emphasize that interpreting
our results within a halo model context requires calibration of how
spectroscopic membership rates relate to halo membership, and that this
relation clearly depends on the adopted halo definition.

\section*{Acknowledgements}

The authors would like to thank August Evrard and Bhuvnesh Jain for useful comments
on an early draft of this work.

This work was supported in part by the U.S. Department of Energy
contract to SLAC no. DE-AC02- 76SF00515, by the National Science
Foundation under NSF-AST-1211838, and by Stanford University, through
a Stanford Graduate Fellowship to RMR.

Funding for SDSS-III has been provided by the Alfred P. Sloan
Foundation, the Participating Institutions, the National Science
Foundation, and the U.S. Department of Energy Office of Science. The
SDSS-III web site is http://www.sdss3.org/.

SDSS-III is managed by the Astrophysical Research Consortium for the
Participating Institutions of the SDSS-III Collaboration including the
University of Arizona, the Brazilian Participation Group, Brookhaven
National Laboratory, Carnegie Mellon University, University of
Florida, the French Participation Group, the German Participation
Group, Harvard University, the Instituto de Astrofisica de Canarias,
the Michigan State/Notre Dame/JINA Participation Group, Johns Hopkins
University, Lawrence Berkeley National Laboratory, Max Planck
Institute for Astrophysics, Max Planck Institute for Extraterrestrial
Physics, New Mexico State University, New York University, Ohio State
University, Pennsylvania State University, University of Portsmouth,
Princeton University, the Spanish Participation Group, University of
Tokyo, University of Utah, Vanderbilt University, University of
Virginia, University of Washington, and Yale University.

GAMA is a joint European-Australasian project based around a
spectroscopic campaign using the Anglo-Australian Telescope. The GAMA
input catalogue is based on data taken from the Sloan Digital Sky
Survey and the UKIRT Infrared Deep Sky Survey. Complementary imaging
of the GAMA regions is being obtained by a number of independent
survey programs including GALEX MIS, VST KiDS, VISTA VIKING, WISE,
Herschel-ATLAS, GMRT and ASKAP providing UV to radio coverage. GAMA is
funded by the STFC (UK), the ARC (Australia), the AAO, and the
participating institutions. The GAMA website is
http://www.gama-survey.org/.

\newcommand\AAA{{A\& A}}
\newcommand\PhysRep{{Physics Reports}}
\newcommand\apj{{ApJ}}
\newcommand\PhysRevD{ {Phys. Rev. D}} 
\newcommand\prd{ {Phys. Rev. D}} 
\newcommand\PhysRevLet[3]{ {Phys. Rev. Letters} }
\newcommand\mnras{{MNRAS}}
\newcommand\PhysLet{{Physics Letters}}
\newcommand\AJ{{AJ}}
\newcommand\aj{{AJ}}
\newcommand\aap{ {A \& A}}
\newcommand\apjl{{ApJ Letters}}
\newcommand\apjs{{ApJ Supplement}}
\newcommand\aph{astro-ph/}
\newcommand\AREVAA{{Ann. Rev. A.\& A.}}
\newcommand{\pasj}{{PASJ}}
\newcommand{\jcap}{{JCAP}}

\bibliographystyle{mn2e}
\bibliography{rm4bib}

\appendix

%%%%%%%%%%%%%%%%%%%%%%%%%%%%%%%%%%%%%%%%
%%%%%%%%%%%%%%%%%%%%%%%%%%%%%%%%%%%%%%%%
%%%%%%%%%%%%%%%%%%%%%%%%%%%%%%%%%%%%%%%%
%%%%%%%%%%%%%%%%%%%%%%%%%%%%%%%%%%%%%%%%
%%%%%%%%%%%%%%%%%%%%%%%%%%%%%%%%%%%%%%%%
%%%%%%%%%%%%%%%%%%%%%%%%%%%%%%%%%%%%%%%%
%%%%%%%%%%%%%%%%%%%%%%%%%%%%%%%%%%%%%%%%
%%%%%%%%%%%%%%%%%%%%%%%%%%%%%%%%%%%%%%%%
%%%%%%%%%%%%%%%%%%%%%%%%%%%%%%%%%%%%%%%%
%%%%%%%%%%%%%%%%%%%%%%%%%%%%%%%%%%%%%%%%
%%%%%%%%%%%%%%%%%%%%%%%%%%%%%%%%%%%%%%%%
%%%%%%%%%%%%%%%%%%%%%%%%%%%%%%%%%%%%%%%%

\section{Photometric Noise Bias in $\chi^2$}
\label{app:chibias}

%%%%%%%%%%%%%%%%%%%%%%%%%%%%%%%%%%%%%%%%
%%%%%%%%%%%%%%%%%%%%%%%%%%%%%%%%%%%%%%%%

\begin{figure*}
\hspace{-12pt} \includegraphics[width=90mm]{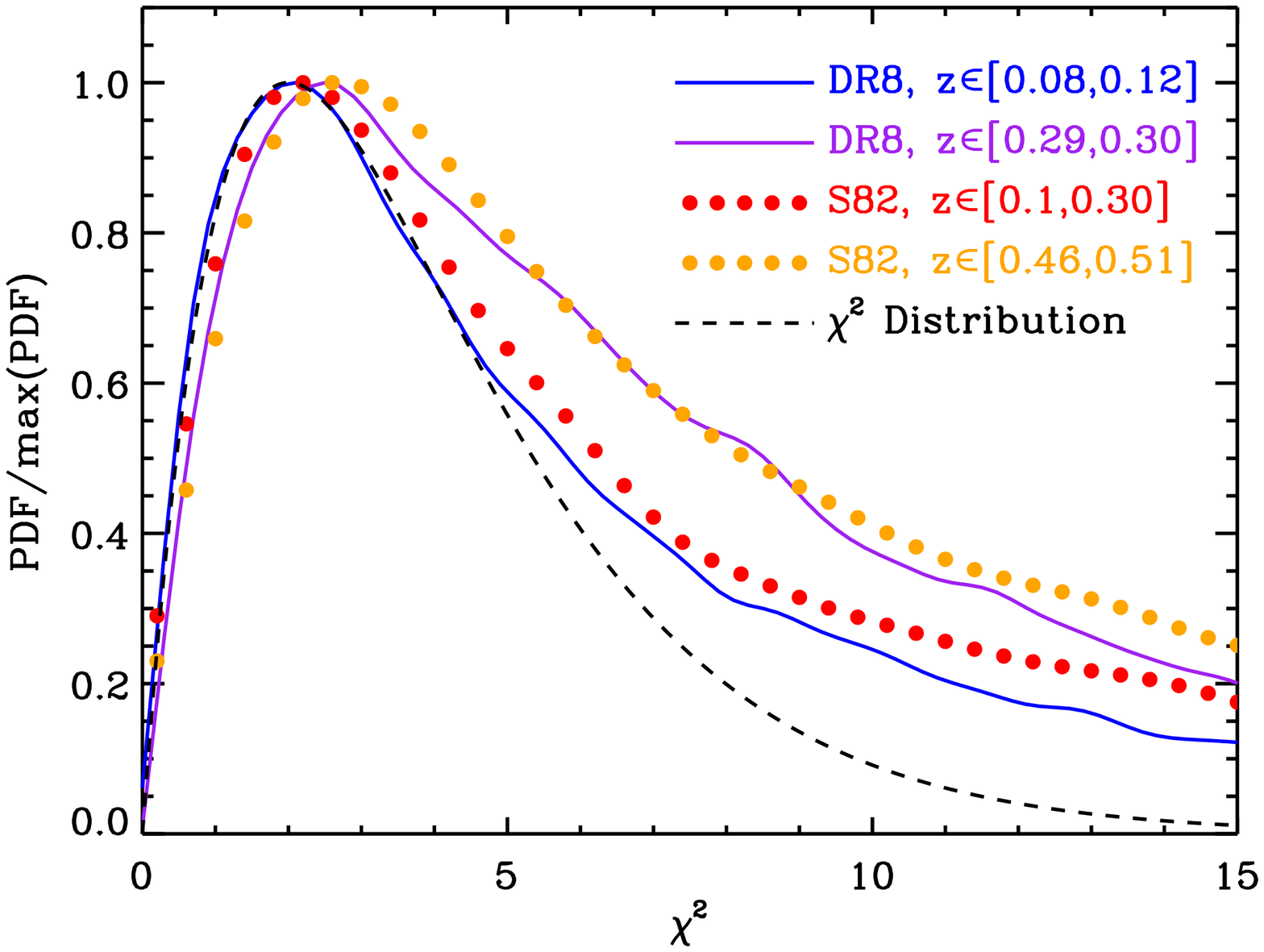}
\hspace{-12pt} \includegraphics[width=90mm]{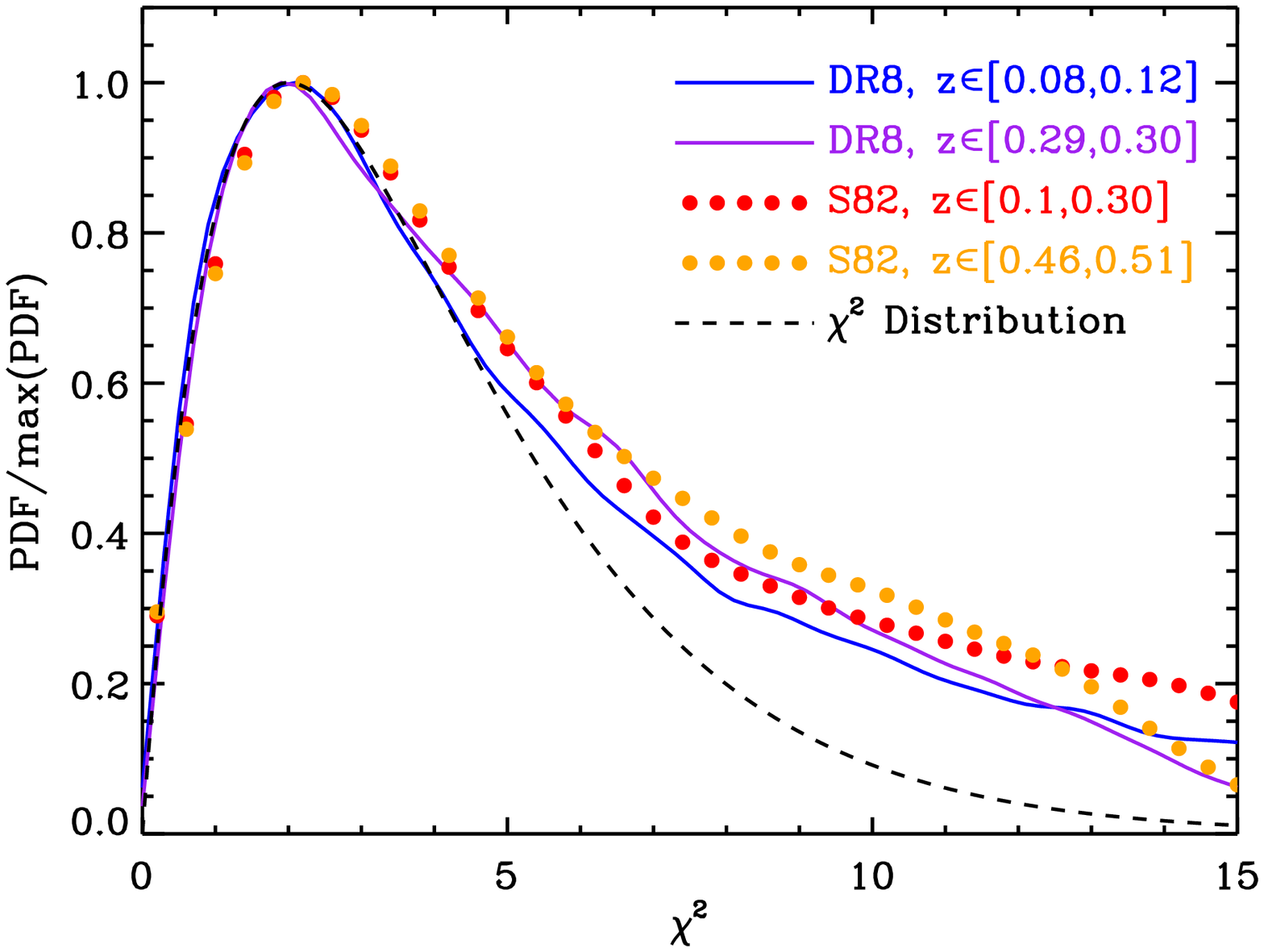}
\caption{
{\bf Left panel:} Comparison of the $\chi^2$ distribution of all galaxies with a membership
probability $\pmem \geq 1\%$ for different data sets and different redshift bins, as labelled.  S82 refers
to \redmapper\ clusters in stripe 82 data.    {\bf Right panel: } As left panel, but we have rescaled
the $\chi^2$ values for the DR8 and S82 high redshift bins as per equation \ref{eq:sdef}
by $s=0.25$.
}
\label{fig:chisqr_pdf_comp}
\end{figure*}

%%%%%%%%%%%%%%%%%%%%%%%%%%%%%%%%%%%%%%%%
%%%%%%%%%%%%%%%%%%%%%%%%%%%%%%%%%%%%%%%%

It was mentioned in section~\ref{sec:pred} that we have discovered that the $\chi^2$
values obtained by \redmapper\ suffer from photometric noise bias. 
The evidence for this bias is shown in Figure~\ref{fig:chisqr_pdf_comp},
where we show the $\chi^2$ distribution of all photometric galaxies with membership
probability $\pmem\geq 1\%$ in the vicinity of \redmapper\ clusters (within
a radius $R_\lambda$) for a variety of clusters in different redshift
bins, as labelled.  In the legend, DR8 refers to \redmapper\ galaxy
clusters in DR8, while S82 refers to \redmapper\ galaxy clusters
in Stripe 82.   

It should be noted that the S82 \redmapper\ catalog
does not use $u$ band data, so the raw $\chi^2$ value from the S82 data is not
directly comparable with that from DR8.  We overplot the two by using density
matching to relate the S82 $\chi^2$ values to the equivalent DR8 $\chi^2$ values.
Given a galaxy with
a $\chi^2$ value $\chi^2_3$ in stripe 82 (the subscript is the number of colors), the corresponding $\chi^2$ value in DR8
will be $\chi^2_4$, selected so as to match the cumulative distribution function of the respective $\chi^2$ distributions.
That is, $\chi^2_4(\chi^2_3)$ is defined via
\be
\int_0^{\chi^2_3} dx\ \rho(x|3) = \int_0^{\chi^2_4} dx\ \rho(x|4).
\ee
This mapping allows us to rescale the $\chi^2$ value for every stripe 82 galaxy into its DR8 equivalent.

We see that the low z and high z DR8 clusters exhibit different $\chi^2$ distributions (blue line vs. purple line),
which could in principle be due to galaxy/cluster evolution.  We see, however, that the distribution of $\chi^2$
values for S82 clusters over the range $z\in[0.1,0.3]$ is identical to the z=0.1 DR8 distribution rather than the z=0.3
DR8 distribution.    Evidently, the difference in the distribution of $\chi^2$ values between the low and high redshift
DR8 samples is not intrinsic evolution, but rather increased photometric noise in the high redshift
DR8 data.  This is confirmed by selecting a stripe 82 redshift bin ($z\in[0.46,0.51]$) for which the median photometric
noise of the cluster galaxies is equal to that of the $z=0.3$ DR8 galaxy sample.  We see that these two distributions (orange
points vs purple line) are identical.

We parameterize the noise bias in $\chi^2$ via a factor $s$ which rescales the observed $\chi^2$ to its correct
value via equation~\ref{eq:s}.  Evidently, the factor $s$ must be $s\approx 0$ for well measured galaxies,
but $s > 0$ for noisy galaxies.  The question is: what does ``noisy'' mean?
Since we are interested
in red-sequence galaxies, the obvious answer is that the rescaling must become necessary when
the observed width of the red-sequence becomes dominated by photometric errors rather than by
its intrinsic width.  Thus, if $\sigma_\mathrm{obs}$ is the photometric error in the galaxy color, and
$\sigma_\mathrm{int}$ is the intrinsic width of the red-sequence in that galaxy color, we expect
that the rescaling factor $s$ will take the form
\be
s = s_\mathrm{max} \frac{\sigma_\mathrm{obs}^2}{\sigma_\mathrm{obs}^2+\sigma_\mathrm{int}^2} \label{eq:s1}
\ee
where $s_\mathrm{max}$ is the maximum value of the rescaling parameter.

In practice, our photometric errors and red-sequence width are multi-dimensional, which
requires a multi-dimensional generalization of equation \ref{eq:s1}.   We make the ansatz
\be
s = \frac{1}{4}\sref \mathrm{Tr}\left( \bC_\mathrm{tot}^{-1}\bC_\mathrm{obs} \right) \label{eq:s}
\ee
where $\bC_\mathrm{tot}$ is the total covariance matrix $\bC_\mathrm{tot}=\bC_\mathrm{int}+\bC_\mathrm{obs}$,
and $\bC_\mathrm{int}$ is the covariance matrix describing the intrinsic scatter of the red-sequence.  The $1/4$
prefactor accounts for the dimensionality of the covariance matrix.  For diagonal matrices, we'd expect
$s \leq \sref$, with $s \rightarrow \sref$ in the limit of very large photometric errors.  In practice, we find
that $s_0$ is roughly equal to the maximum value for $s$ observed in our galaxies, but it is not a strict upper bound.
Note that since $\bC_\mathrm{int}$ is a function of redshift, we expect $\sref$ to have some mild redshift
dependence.

To compute our best fit model for $\sref$, we proceed as follows.  First,  we rescale our 
data as per equations \ref{eq:sdef} and \ref{eq:s}.  The cluster member galaxies are then binned
to arrive at an empirical estimate of the rescaled $\chi^2$ distribution.   This distribution
is expected to match a $\chi^2$ distribution, so we construct a cost function 
$E(\sref)$ defined as the total square deviation between our empirical
estimate and our model prediction (which is properly integrated over each $\chi^2$ bin).   
Our best fit model for $\sref$ is that which minimizes our
cost function.  In order to ensure
that the fit is done over a region that is well described by a $\chi^2$ distribution, we only fit
the region $\chi^2_s \leq 4$.
Further, we allow $\sref$ to be redshift dependent, with $\sref(z)$ being parameterized using spline interpolation.
The model parameters are the values on the spline nodes, for which we set $z=0.08$, $z=0.28$, and $z=0.55$.
We find that three nodes are sufficient to accurately model the full $z\in[0.08,0.55]$ redshift range.

\begin{figure*}
\hspace{-12pt} \includegraphics[width=90mm]{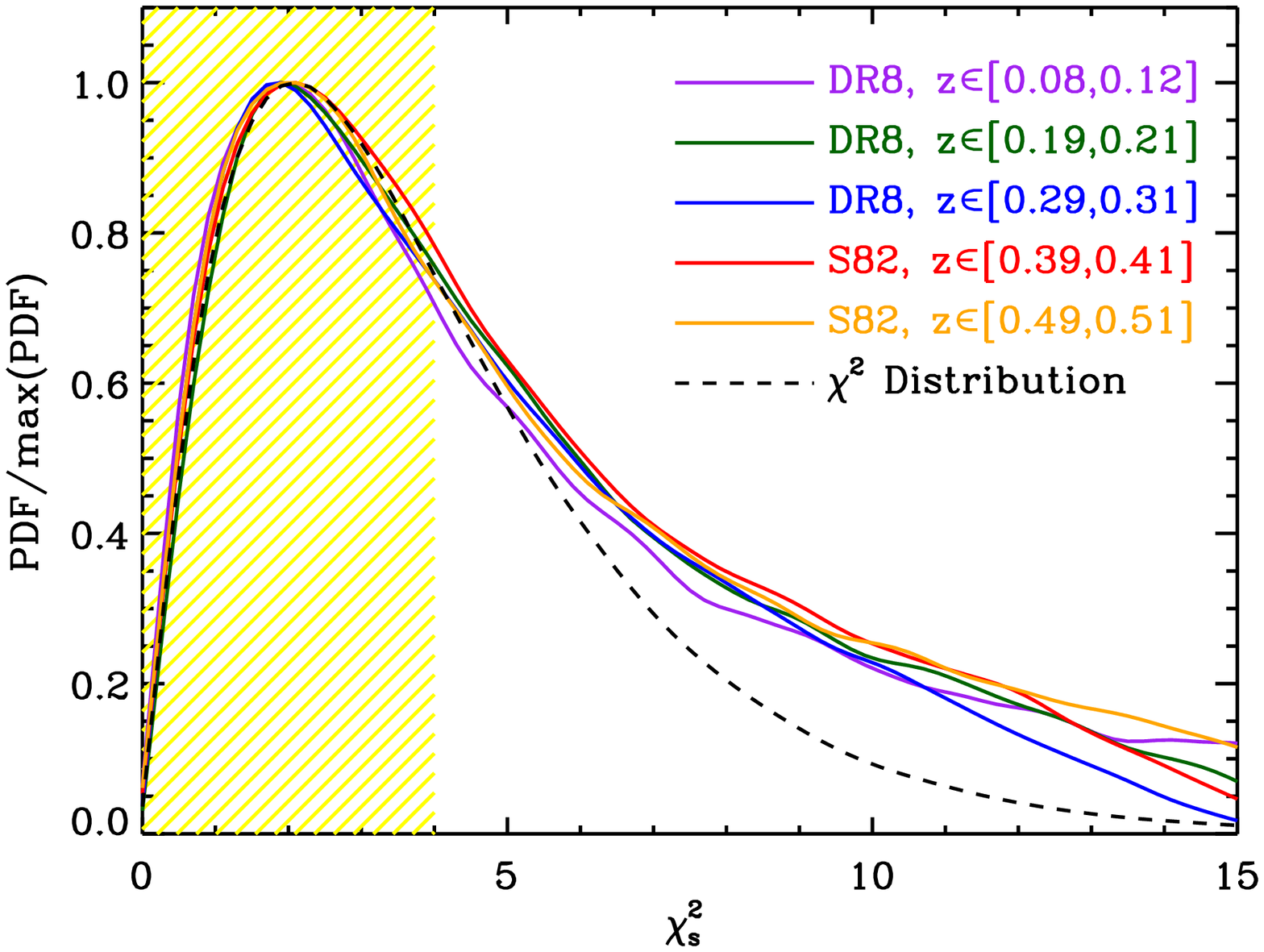}
\hspace{-12pt} \includegraphics[width=90mm]{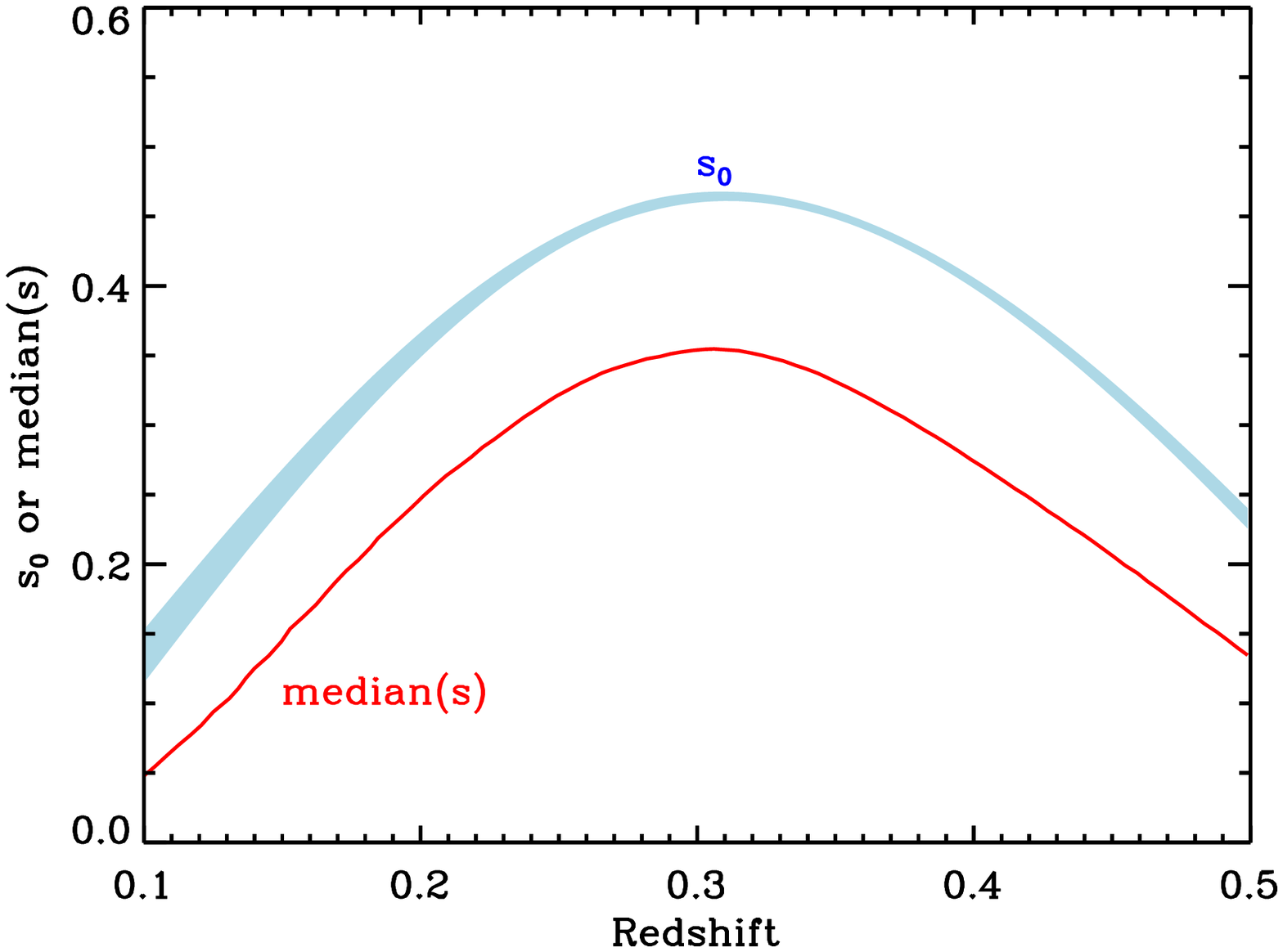}
\caption{
{\bf Left panel:} Distribution of the rescaled $\chi^2$ values of cluster galaxies in DR8 
as per equations \ref{eq:sdef} and \ref{eq:s} using our best fit model for $\sref(z)$
for a variety of different redshifts, as labelled.  The hatched yellow region denotes the
range of $\chi^2$ values used in fitting for $\sref$.  {\bf Right panel:} The 68\% confidence
band for $\sref(z)$, as well as the median value of $s$ for all DR8 cluster galaxies
as a function of redshift.
}
\label{ref:spectest_chisqr_pdf}
\end{figure*}

The left panel in Figure~\ref{ref:spectest_chisqr_pdf} shows the distribution of rescaled $\chi^2$ values obtained using our
best fit model for $\sref(z)$. It is immediately apparent that our model does a good job of accounting for the bias
introduced by photometric noise in our measurement.  Importantly, the distribution of rescaled $\chi^2$ values
is universal not only over the region $\chi^2 \leq \chi^2_\mathrm{cut}$, but across the entire range of $\chi^2$
values that we probe.  

%%%%%%%%%%%%%%%%%%%%%%%%%%%%%%%%%%%%%%%%
%%%%%%%%%%%%%%%%%%%%%%%%%%%%%%%%%%%%%%%%

%%%%%%%%%%%%%%%%%%%%%%%%%%%%%%%%%%%%%%%%
%%%%%%%%%%%%%%%%%%%%%%%%%%%%%%%%%%%%%%%%

The right panel shows our best fit model for $\sref(z)$, as well as the median shift
$s(z)$ for cluster galaxies at redshift $z$. For $\sref$, we show the 68\% confidence
band at each redshift, as determined by bootstrap resampling the cluster member galaxy
catalog 100 times and then recomputing the best fit node values for each realization.   
In addition, we also show the median $s$ value of our cluster galaxies as a function of redshift.

%%%%%%%%%%%%%%%%%%%%%%%%%%%%%%%%%%%%%%%%
%%%%%%%%%%%%%%%%%%%%%%%%%%%%%%%%%%%%%%%%
%%%%%%%%%%%%%%%%%%%%%%%%%%%%%%%%%%%%%%%%
%%%%%%%%%%%%%%%%%%%%%%%%%%%%%%%%%%%%%%%%
%%%%%%%%%%%%%%%%%%%%%%%%%%%%%%%%%%%%%%%%
%%%%%%%%%%%%%%%%%%%%%%%%%%%%%%%%%%%%%%%%
%%%%%%%%%%%%%%%%%%%%%%%%%%%%%%%%%%%%%%%%
%%%%%%%%%%%%%%%%%%%%%%%%%%%%%%%%%%%%%%%%
%%%%%%%%%%%%%%%%%%%%%%%%%%%%%%%%%%%%%%%%
%%%%%%%%%%%%%%%%%%%%%%%%%%%%%%%%%%%%%%%%
%%%%%%%%%%%%%%%%%%%%%%%%%%%%%%%%%%%%%%%%
%%%%%%%%%%%%%%%%%%%%%%%%%%%%%%%%%%%%%%%%

\section{Updates to the redMaPPer Algorithm}
\label{app:updates}

The analysis in this paper relies on the \redmapper\ catalog obtained using the \redmapper\
code version 5.10, which includes a variety of updates and upgrades to version 5.2, used in Paper I.  
We summarize these changes here, and make the \redmapper\ v5.10 catalog publicly available
with this work.

In addition to fixing assorted minor bug fixes, the changes to the \redmapper\ algorithm between v5.2 and v5.10 are:

1) {\bf Clusters are selected directly on the number of detected cluster
  galaxies.}  As with \redmapper{} v5.2, our detection threshold is set to 20
detected cluster galaxies.  In a region
where no galaxies are masked (no star holes and where we are complete to the
luminosity threshold), this is equivalent to a $\lambda > 20$ threshold.
However, if part of the cluster is masked, 20 galaxy detections must
necessarily correspond to a richness threshold larger than 20.  In \redmapper{}
v5.2, we set this threshold as a simple function of redshift, based on the
average depth of the survey and ignoring the effects of star holes and
boundaries.  In v5.10, we directly set a threshold for each individual cluster
to ensure 20 galaxies are detected.  For fairly uniform surveys such as SDSS,
this change has a very small impact on cluster selection.  However, future
surveys such as the Dark Energy Survey (DES) have much larger depth
variations.  In the interest of making our algorithm more generally applicable,
we have applied this update when running on SDSS data as well.

2) {\bf All cuts used in defining richness are treated as smooth rather than sharp cuts.}
In \redmapper\ v5.2, in the absence of masking the cluster richness
was defined via
\be
\lambda = \sum p_i
\ee
where $p_i$ is the membership probability of galaxy $i$.  The sum was restricted to galaxies brighter than $0.2L_*$ and 
within a radial separation $R \leq R(\lambda)$, where $R(\lambda)$ is a richness-dependent aperture.  Consequently, the above
equation can be rewritten as
\be
\lambda = \sum p_i \theta_i^L\theta_i^R
\ee
where the sum is now over {\it all} galaxies, and $\theta_i^L$ and $\theta_i^R$
are luminosity and radius dependent weights, which are top-hat functions for
\redmapper{} v5.2.  It is clear from this that the richness definition is inherently unstable: there are always cases of galaxies that are just over or just 
under $0.2L_*$ in luminosity, and/or just inside or just outside the radius
$R\leq R(\lambda)$.  Therefore, small changes to either ``edge''
can result in macroscopic changes to the richness.

To overcome this difficulty, we now utilize soft cut-off weights.  Specifically, we set
\bea
\theta_i^L & = & \half \left[ 1+ \erf\left( \frac{m_{\rm max} - m_i}{\sigma_i} \right)\right] \\
\theta_i^R & = & \half \left[ 1 + \erf \left( \frac{R(\lambda) - R}{\sigma_R} \right) \right],
\eea
where $m_{\rm max}$ is the magnitude corresponding to the $0.2L_*$ luminosity
threshold or the survey limiting magnitude (whichever is brighter), $\sigma_i$ is the photometric error of galaxy $i$,
and $\sigma_R=0.05\ \hMpc$.  This has a small impact on the richness of most galaxy clusters, while making the richness stable for those
clusters with galaxies just inside or just outside our fiducial boundaries.

3) {\bf Cluster galaxy mask-fractions are now estimated taking into account the local survey depth.}
Galaxies are now selected based on the local depth of the SDSS imaging, which is estimated ``on the fly'' on a cluster
by cluster basis. In Rykoff et al. (in prep), we describe a method for
estimating the depth of a photometric survey based on the galaxy catalog. 
The idea is simple: given the effective sky noise, and assuming Poisson errors
in the photon counts, one can derive a two-free parameter model that relates the magnitude of a source to its error.  These two
free parameters are the effective exposure time and the $10\sigma$ limiting magnitude of the image. 
We estimate the depth in each cluster field by selecting all galaxies within a $2.5\ \hMpc$ aperture of the cluster center,
and fitting our model to the resulting galaxy data.  This information is utilized when computing mask-fraction corrections.
Specifically, when computing mask fraction corrections, we generate Monte Carlo realizations of the cluster galaxies, and
then perturb their magnitudes in accordance with the local depth to estimate the fraction of the cluster being masked.

4) {\bf Our propagation of the uncertainty introduced by masking into richness errors has been updated to make it significantly more stable.}  
Specifically, in the presence
of masking, the cluster richness is estimated via
\be
\lambda(1-C) = \sum p_i  \label{eq:c}
\ee
where $C$ is the fraction of the cluster being masked, which has an associated uncertainty $\sigma_C$.  For details,
see Paper I.   We convert the uncertainty in $C$ into a richness error estimate via
\be
\sigma_\lambda = \left\vert \frac{d\lambda}{dC} \right\vert \sigma_C. \label{eq:var1}
\ee

In \redmapper\ v5.2, the factor $d\lambda/dC$ was estimated numerically using a finite difference method.  However, we found
this procedure to be numerically noisy.  Here, we rely on an alternative method for computing $d\lambda/dC$.  Specifically,
using the fact that the membership probability of a galaxy is $p=\lambda u/(\lambda u+b)$, and taking the differential of equation~\ref{eq:c},
we arrive at
\be
dC = \sum \left ( \frac{u}{\lambda u +b} \right ) ^2 u d\lambda,
\ee
or simply
\be
\frac{d\ln \lambda}{dC} = \frac{1}{\lambda}\frac{1}{\sum p^2}
\ee
where the sum is over the detected galaxies, and is evaluated using a {\it fixed} metric aperture.   We plug this into equation~\ref{eq:var1}
to get the error in the cluster richness given the fixed metric aperture.
Note, however, that the aperture used for richness estimation itself depends on
richness.  Therefore, an increased richness leads to a larger aperture, which
in turn leads to an even larger richness estimate.
An increase in richness $d\ln \lambda_0$ at fixed aperture will increase the corresponding aperture via,
\be
d \ln R_1 = \beta d\ln \lambda_0,
\ee
with the factor of $\beta$ coming from the relation between cluster richness and the cluster aperture, ($R(\lambda)\propto \lambda^{\beta}$, see Paper I).  
If the cluster richness profile is such that $\lambda(R) \propto R^\gamma$, then the above aperture change will
further increase the richness by 
\be
d\ln \lambda_ 1 = \gamma d\ln R_1 = \beta \gamma d\ln \lambda_0.
\ee
The net richness change is then: 
\bea
d\ln \lambda & = & \sum_i d\ln \lambda_i \\
	& = & d\ln \lambda_0 \sum_{i=0}^{\infty} (\beta\gamma)^i \\
	& = & \frac{d\ln \lambda_0}{1-\beta\gamma}.
\eea
Consequently, our final estimate for the richness error due to masking is given
by
\be
\sigma_\lambda = \frac{1}{1-\beta\gamma}\frac{1}{\sum p^2} \sigma_C.
\ee
Now $\sigma_C$ is estimated precisely as in Paper I; the only difference between our current v5.10 analysis
and that described in Paper I is the prefactor in front of $C$ in the equation above.
We note that the value $\beta=0.2$ is set by the radius--richness relation in Paper I, while
the factor $\gamma$ is the local slope of the richness profile of galaxy clusters.  We measure
this by cluster stacking, finding $\gamma=0.6$, which we adopt as our fiducial value.  

We have explicitly verified that our new estimates for the richness error estimates are, by and large,
in agreement with those in Paper I, except the new estimates are significantly more accurate because
of reduced numerical noise relative to the finite difference method employed in Paper I.

%%%%%%%%%%%%%%%%%%%%%%%%%%%%%%%%%%%%%%%%
%%%%%%%%%%%%%%%%%%%%%%%%%%%%%%%%%%%%%%%%
%%%%%%%%%%%%%%%%%%%%%%%%%%%%%%%%%%%%%%%%
%%%%%%%%%%%%%%%%%%%%%%%%%%%%%%%%%%%%%%%%
%%%%%%%%%%%%%%%%%%%%%%%%%%%%%%%%%%%%%%%%
%%%%%%%%%%%%%%%%%%%%%%%%%%%%%%%%%%%%%%%%
%%%%%%%%%%%%%%%%%%%%%%%%%%%%%%%%%%%%%%%%
%%%%%%%%%%%%%%%%%%%%%%%%%%%%%%%%%%%%%%%%
%%%%%%%%%%%%%%%%%%%%%%%%%%%%%%%%%%%%%%%%
%%%%%%%%%%%%%%%%%%%%%%%%%%%%%%%%%%%%%%%%
%%%%%%%%%%%%%%%%%%%%%%%%%%%%%%%%%%%%%%%%
%%%%%%%%%%%%%%%%%%%%%%%%%%%%%%%%%%%%%%%%

\section{Description of Columns in the DR8 Cluster Catalog}
\label{app:catalog}

The full \redmapper{} DR8 cluster and member catalogs are available at {\tt
  http://risa.stanford.edu/redmapper/} in FITS format, and from the online
journal in machine-readable formats.  A summary of the cluster catalog
information is given in Table~\ref{tab:catkey}.  A summary of the member
information is given in Table~\ref{tab:memkey}.

\begin{table*}
\centering
\begin{minipage}{150mm}
\caption{\redmapper{} DR8 Cluster Catalog Format}
\begin{tabular}{llcl}
Column & Name & Format & Description \\
1 & ID & I7 & \redmapper{} Cluster Identification Number\\
2 & NAME & A20 & \redmapper{} Cluster Name\\
3 & RA & F12.7 & Right ascension in decimal degrees (J2000)\\
4 & DEC & F12.7 & Declination in decimal degrees (J2000)\\
5 & Z\_LAMBDA & F6.4 & Cluster \photoz $\zlambda$\\
6 & Z\_LAMBDA\_ERR & F6.4 & Gaussian error estimate for $\zlambda$\\
7 & LAMBDA & F6.2 & Richness estimate $\lambda$\\
8 & LAMBDA\_ERR & F6.2 & Gaussian error estimate for $\lambda$\\
9 & S & F6.3 & Richness scale factor (see Eqn.~\ref{eqn:sz})\\
10 & Z\_SPEC & F8.5 & SDSS spectroscopic redshift for most likely center (-1.0 if not available)\\
11 & OBJID & I20 & SDSS DR8 CAS object identifier\\
12 & IMAG & F6.3 & $i$-band cmodel magnitude for most likely central galaxy (dereddened)\\
13 & IMAG\_ERR & F6.3 & error on $i$-band cmodel magnitude\\
14 & MODEL\_MAG\_U & F6.3 & $u$ model magnitude for most likely central galaxy
(dereddened)\\
15 & MODEL\_MAGERR\_U & F6.3 & error on $u$ model magnitude\\
16 & MODEL\_MAG\_G & F6.3 & $g$ model magnitude for most likely central galaxy
(dereddened)\\
17 & MODEL\_MAGERR\_G & F6.3 & error on $g$ model magnitude\\
18 & MODEL\_MAG\_R & F6.3 & $r$ model magnitude for most likely central galaxy
(dereddened)\\
19 & MODEL\_MAGERR\_R & F6.3 & error on $r$ model magnitude\\
20 & MODEL\_MAG\_I & F6.3 & $i$ model magnitude for most likely central galaxy
(dereddened)\\
21 & MODEL\_MAGERR\_I & F6.3 & error on $i$ model magnitude\\
22 & MODEL\_MAG\_Z & F6.3 & $z$ model magnitude for most likely central galaxy
(dereddened)\\
23 & MODEL\_MAGERR\_Z & F6.3 & error on $z$ model magnitude\\
24 & ILUM & F7.3 & Total membership-weighted $i$-band luminosity (units of
$L_*$)\\
25 & P\_CEN[0] & E9.3 & Centering probability $\Pcen$ for most likely central\\
26 & RA\_CEN[0] & F12.7 & R.A. for most likely central\\
27 & DEC\_CEN[0] & F12.7 & Decl. for most likely central\\
28 & ID\_CEN[0] & I20 & DR8 CAS object identifier for most likely central\\
29-32 & \_CEN[1] & & $\Pcen$, R.A., Decl., and ID for second most likely central\\
33-36 & \_CEN[2] & & $\Pcen$, R.A., Decl., and ID for third most likely central\\
37-40 & \_CEN[3] & & $\Pcen$, R.A., Decl., and ID for fourth most likely central\\
41-44 & \_CEN[4] & & $\Pcen$, R.A., Decl., and ID for fifth most likely
central\\
45-65 & PZBINS & F7.4 & Redshift points at which $P(z)$ is evaluated\\
66-86 & PZ & E10.3 & $P(z)$ evaluated at redshift points given by PZBINS\\
\end{tabular}
\label{tab:catkey}
\tablecomments{This table is presented in its entirety in the online edition of
  the journal, and at {\tt
    http://risa.stanford.edu/redmapper}.}
\end{minipage}
\end{table*}

\begin{table*}
\centering
\begin{minipage}{150mm}
\caption{\redmapper{} DR8 Member Catalog Format}
\begin{tabular}{llcl}
Column & Name & Format & Description \\
1 & ID & I7 & \redmapper{} Cluster Identification Number\\
2 & RA & F12.7 & Right ascension in decimal degrees (J2000)\\
3 & DEC & F12.7 & Declination in decimal degrees (J2000)\\
4 & R & F5.3 & Distance from cluster center ($h^{-1}\,\mathrm{Mpc}$)\\
5 & P & F5.3 & Membership probability\\
6 & P\_SPEC & F5.3 & Spectroscopic calibrated membership probability\\
7 & P\_FREE & F5.3 & Probability that member is not a member of a higher-ranked
cluster\\
8 & THETA\_I & F5.3 & Luminosity ($i$-band) weight\\
9 & THETA\_R & F5.3 & Radial weight\\
10 & IMAG & F6.3 & $i$-band cmodel magnitude (dereddened)\\
11 & IMAG\_ERR & F6.3 & error on $i$-band cmodel magnitude\\
12 & MODEL\_MAG\_U & F6.3 & $u$ model magnitude 
(dereddened)\\
13 & MODEL\_MAGERR\_U & F6.3 & error on $u$ model magnitude\\
14 & MODEL\_MAG\_G & F6.3 & $g$ model magnitude
(dereddened)\\
15 & MODEL\_MAGERR\_G & F6.3 & error on $g$ model magnitude\\
16 & MODEL\_MAG\_R & F6.3 & $r$ model magnitude
(dereddened)\\
17 & MODEL\_MAGERR\_R & F6.3 & error on $r$ model magnitude\\
18 & MODEL\_MAG\_I & F6.3 & $i$ model magnitude
(dereddened)\\
19 & MODEL\_MAGERR\_I & F6.3 & error on $i$ model magnitude\\
20 & MODEL\_MAG\_Z & F6.3 & $z$ model magnitude
(dereddened)\\
21 & MODEL\_MAGERR\_Z & F6.3 & error on $z$ model magnitude\\
22 & Z\_SPEC & F8.5 & SDSS spectroscopic redshift (-1.0 if not available)\\
23 & OBJID & I20 &  SDSS DR8 CAS object identifier\\
\end{tabular}
\tablecomments{The photometric probability $P$ is the original \redmapper\ photometric
membership probability, while the probability $P_{\mathrm free}$ is the probability that
the galaxy does not belong to a previous cluster in the percolation.  Thus, the total
membership probability is $P\times P_{\rm free}$.  Finally, the probability $P_{\mathrm spec}$
is the probability that a galaxy is a spectroscopic member, estimated as detailed in this work.}
\end{minipage}
\label{tab:memkey}
\end{table*}

\label{lastpage}

\end{document}